\documentclass[fleqn,10pt]{wlscirep}
\usepackage{color,soul}
\usepackage{csquotes}
\usepackage{nameref}
\usepackage{xr}
\usepackage{prettyref}

\title{Snap-through transition of graphene membranes for memcapacitor
applications: A~combined study using MD, DFT and elasticity theory}

\author[1,2,*]{Ruslan~D.~Yamaletdinov}

\author[3,4]{Oleg~V.~Ivakhnenko}

\author[1,2]{Olga~V.~Sedelnikova}

\author[3,4]{Sergey~N.~Shevchenko}

\author[1,5,$\dagger$]{Yuriy~V.~Pershin}

\affil[1]{Nikolaev Institute of Inorganic Chemistry SB RAS, Novosibirsk 630090, Russia}
\affil[2]{Novosibirsk State University, Novosibirsk 630090, Russia}
\affil[3]{B.~I. Verkin Institute for Low Temperature Physics and Engineering, Kharkov 61103, Ukraine}
\affil[4]{V.~N. Karazin Kharkov National University, Kharkov 61022, Ukraine}
\affil[5]{Department of Physics and Astronomy, University of South Carolina, Columbia,
South Carolina 29208, USA}
\affil[*]{yamaletdinov@niic.nsc.ru}
\affil[$\dagger$]{pershin@physics.sc.edu}


\begin{abstract}
Using computational and theoretical approaches, we investigate the
snap-through transition of buckled graphene membranes. Our main interest is
related to the possibility of using the buckled membrane as a plate of
capacitor with memory (memcapacitor). For this purpose, we performed
molecular-dynamics (MD) simulations and elasticity theory calculations of
the up-to-down and down-to-up snap-through transitions for membranes of
several sizes. We have obtained expressions for the threshold switching
forces for both up-to-down and down-to-up transitions. Moreover, the
up-to-down threshold switching force was calculated using the density
functional theory (DFT). Our DFT results are in general agreement with MD
and analytical theory findings. Our systematic approach can be used
for the description of other structures, including nanomechanical and biological ones,
experiencing the snap-through transition.
\end{abstract}

\begin{document}

\flushbottom
\maketitle

\thispagestyle{empty}

\section*{Introduction}

\label{sec1}

Memcapacitors \cite{diventra09a} are an emerging type of circuit elements
with memory whose instantaneous response depends on the internal state and
input signal. Such devices are prospective candidates for applications in
information storage and processing~\cite{traversa14a,pershin15a}
technologies as their states can be manipulated by the applied voltages or
charges and can store information for long intervals
of time. Several possible realizations of memcapacitors were suggested by
using micro-electro-mechanical systems~\cite{pershin11c}, ionic transport~%
\cite{Lai09a}, electronic effects~\cite{martinez09a}, superconducting qubits
\cite{Shevchenko16}, etc. \cite{pershin11a}

Generally, voltage-controlled memcapacitive systems (memcapacitors) are
described by~\cite{diventra09a}
\begin{eqnarray}
q(t) &=&C\left( x,V,t\right) V(t),  \label{Ceq1} \\
\dot{x} &=&f\left( x,V,t\right) , \label{Ceq2}
\end{eqnarray}%
where $q(t)$ is the charge on the capacitor at time $t$, $V(t)$ is the
applied voltage, $C$ is the \textit{memcapacitance} (memory capacitance), $x$
is a set of $n$ internal state variables, and $f$ is a continuous $n$%
-dimensional vector function. In some cases, it is more convenient to
consider charge-controlled memcapacitors~\cite{diventra09a} such that the
charge instead of voltage is considered as input.

\begin{figure}[b]
\vspace{0.6cm}
\centering{\includegraphics[width=0.5\columnwidth]{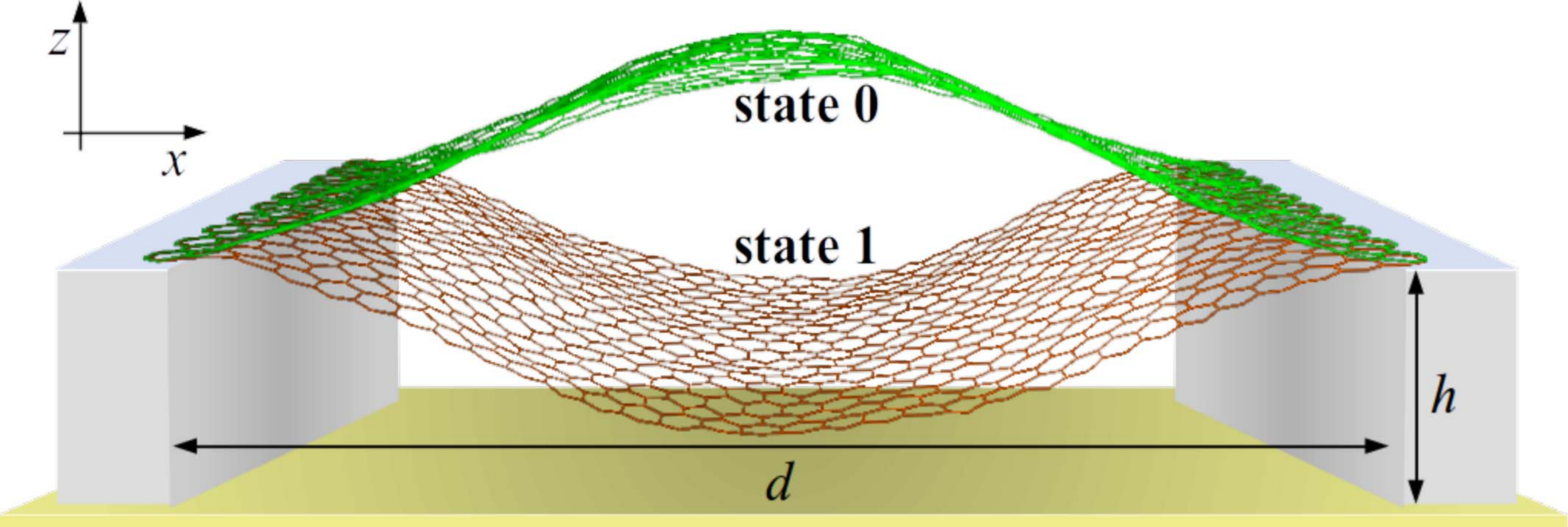}}
\caption{Schematics of the membrane memcapacitor
employing a buckled graphene membrane as its top plate~\protect\cite{pershin11c}.}
\label{fig1}
\end{figure}

Among several possible realizations of memcapacitors, the membrane-based
memcapacitors~\cite{pershin11c} are of significant interest as their
geometry makes them intrinsically suitable for non-volatile storage of
binary information. Indeed, the buckled membrane used as the top capacitor
plate (see Fig. \ref{fig1} for schematics) has two
stable buckled states corresponding to two distinct values of capacitance.
It was suggested~\cite{pershin11c} that the switching between these states
can be performed using the attractive interaction of oppositely charged
capacitor plates. Moreover, it was demonstrated theoretically that simple
circuits of membrane memcapacitors offer an in-memory computing
functionality~\cite{pershin15a}.

In this work, we consider a possible realization of membrane-based
memcapacitor~\cite{pershin11c} employing a single- or multi-layer graphene
membrane~\cite{Lindahl12, Weber14, Lambin14, Benameur15, Davidovikj16} as
its bistable plate (see Fig. \ref{fig1}). Our aim is to understand the basic
physical processes and parameters underlying the snap-through transition of
such membrane including details of the membrane dynamics, threshold forces, etc.
For this purpose, we
perform a combined study using MD, DFT and elasticity theory focusing on a
single-layer graphene membrane with clamped boundary conditions. This choice
of boundary conditions is justified by the typically strong adhesion of
graphene to substrates. Our results extend our prior DFT investigation~%
\cite{Sedelnikova2016} of the up-to-down snap-through transition of
graphene membrane with hinged boundary conditions.

The combination of computational/theoretical methods adds breadth and depth
to our analysis. Using MD simulations, we were able to understand main
features of the membrane dynamics in the presence of an external force and
after the force removal. This understanding has helped us to develop
analytical models that resulted in compact algebraic expressions for the
threshold switching forces. DFT calculations were used to validate MD
results for the up-to-down transition.

This paper is organized as follows. In Sec. "\nameref{sec2}" we investigate the
snap-through transition of graphene membranes using molecular dynamics
simulations. In particular, MD simulations of the up-to-down and down-to-up
transitions are reported in Subsec. "\nameref{sec2b}" and "\nameref{sec2c}", respectively,
while MD simulation details can be found in Supplementary Information (SI) Sec. "MD Simulation details".
The standard elasticity theory is applied to the membrane switching in Subsec.~"\nameref{Sec:Elasticity}".
A phenomenological analytical model of the
snap-through transition is presented in Subsec.~"\nameref{sec3}" and in SI Sec. ”Phenomenological Elasticity Theory”.
Our DFT calculations are summarized in Sec.~"\nameref{Sec:DFTmain}".
The results obtained within different approaches as well as their implications are
discussed in Sec.~"\nameref{Sec:Discussion}".

\vspace{0.2cm}

In this paper, the following notations are used:

\vspace{0.2cm}

$q$ - the charge on capacitor (see Eq. (\ref{Ceq1}))

$V$ - the applied voltage (see Eqs. (\ref{Ceq1}), (\ref{Ceq2}))

$C$ - the (memory) capacitance (see Eq. (\ref{Ceq1}))

$d$ - the distance between fixed sides of membrane (see Fig. \ref{fig1})

$h$ - the distance between the bottom plate and the level of fixed sides (see Fig. \ref{fig1})

$L$ - the membrane length

$w$ - the membrane width

$D= 1.6$ eV - the bending rigidity of graphene

$E_{2D}=340$ N/m - the 2D Young's module

$\zeta$ - the deflection of membrane (see Eq. (\ref{U}))

$\zeta _{\mathrm{c}}$ - the maximum deflection of membrane (see Eq. (\ref%
{U_buckled}))

$\theta _{i}(s)$ - the angle that the membrane makes with the horizontal (see
Eqs. (\ref{eq:theta_s}) and (\ref{eq:theta_ns})),

$s$ - the internal coordinate that changes between $-1/2$ and $1/2$ (see
Eqs. (\ref{eq:theta_s}) and (\ref{eq:theta_ns}))

$A_{i}$ and $c_{i}$ - coefficients (see Eqs. (\ref{eq:theta_s}) and (\ref%
{eq:theta_ns}))

$z_{cm}$ - the center of mass position (see Eq. (\ref{eq22}))

$U_{b}$, $U_{str}$, $U_{ext}$ - the bending, stretching and external potential
contributions to the potential energy of membrane (see Eq.~(\ref{eq:pot_energ}))

$F^\downarrow$ - the up-to-down threshold switching force (see Eqs. (\ref{F_star}), (\ref{eq20}), (\ref{eq888}), and (\ref%
{eq:fsw_nonsym}))

$F^\uparrow$ - the down-to-up threshold switching force (see Eqs. (\ref{Fup}) and (\ref{FZCM}%
))

$\varepsilon_0 $ - the vacuum permittivity

\section{Molecular Dynamics Simulations}\label{sec2}

MD simulations are a well
established modeling tool frequently employed in studies of nanoscale
carbon-based materials~\cite{Yakobson1996, Berber2000, Yao2001, Legoas2003,
Maruyama2003, Lee2006, Shiomi2006, Wang2009, Hu2009, Jiang2009, martins2010,
Neek-Amal2010, Ni2010, Lebedeva2011, Bagri2011, Min2011, Ng2012,
Rajabpour2012, Kalosakas2013, Berdiyorov2014, Kang2015, Yamaletdinov17a}.
We used classical MD simulations to investigate the dynamics of snap-through
transition of buckled graphene membranes. Zigzag graphene nanoribbons (membranes) of two
lengths were considered: the nanoribbon \textbf{\textit{A}}, $L=54$ \AA\ (22
rings), and nanoribbon \textbf{\textit{B}}, $L=103$ \AA\ (42 rings). Both
nanoribbons were of the same width ($w=41$ \AA). In order to implement the
clamped boundary conditions, the first two lines of carbon atoms at shorter
sides were kept fixed. The buckling was realized by changing the distance
between the fixed sides from $L$ to $d<L$. The same external force was
applied to each atom in the downward direction to simulate the attractive
interaction between the plates.

See SI Sec. "MD Simulation details" for the details of MD simulations.

\subsection{Up-to-down transition}

\label{sec2b}

\begin{figure}[tb]
\centering{(a)\includegraphics[width=0.42\columnwidth]{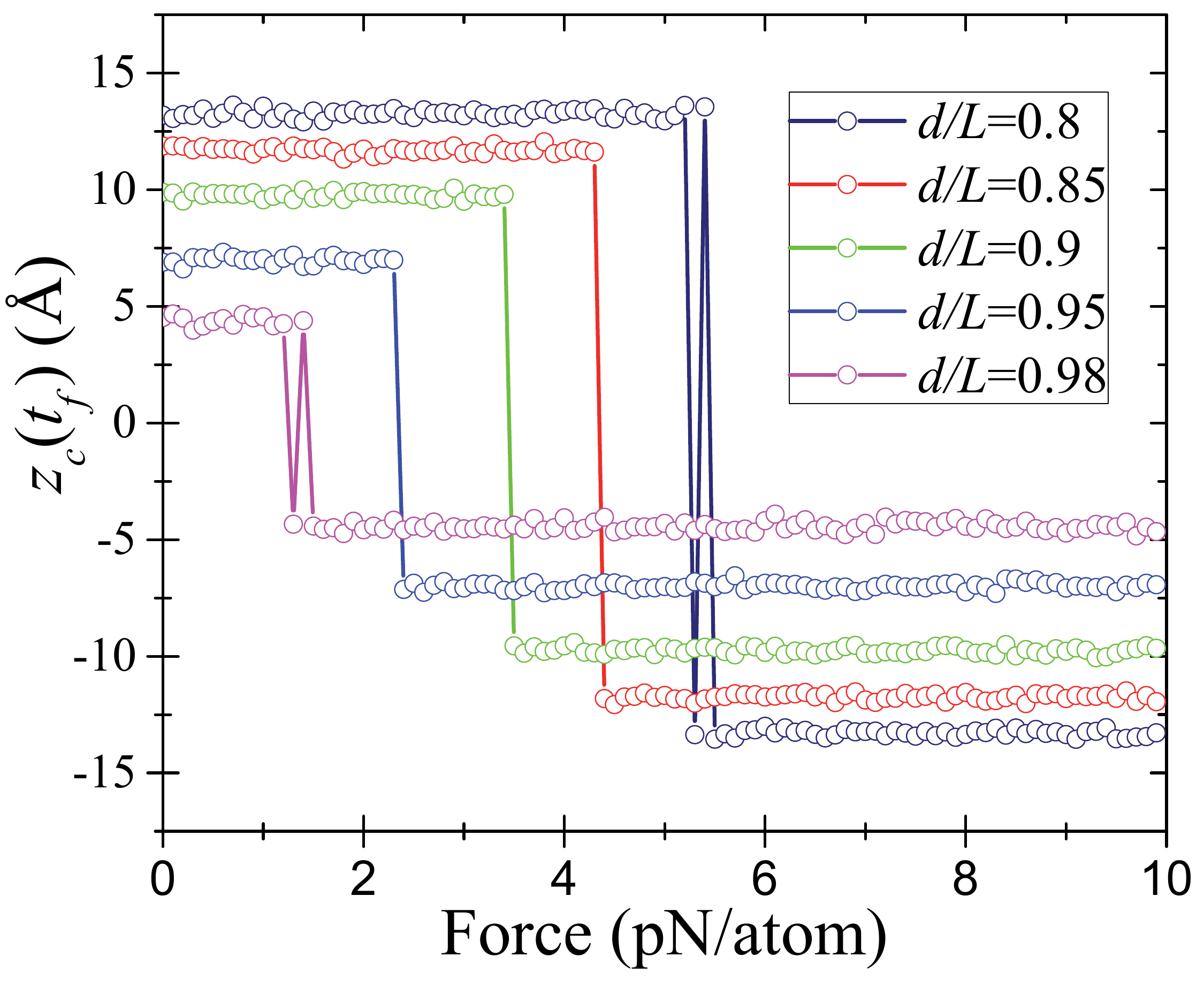} (b)
\includegraphics[width=0.42\columnwidth]{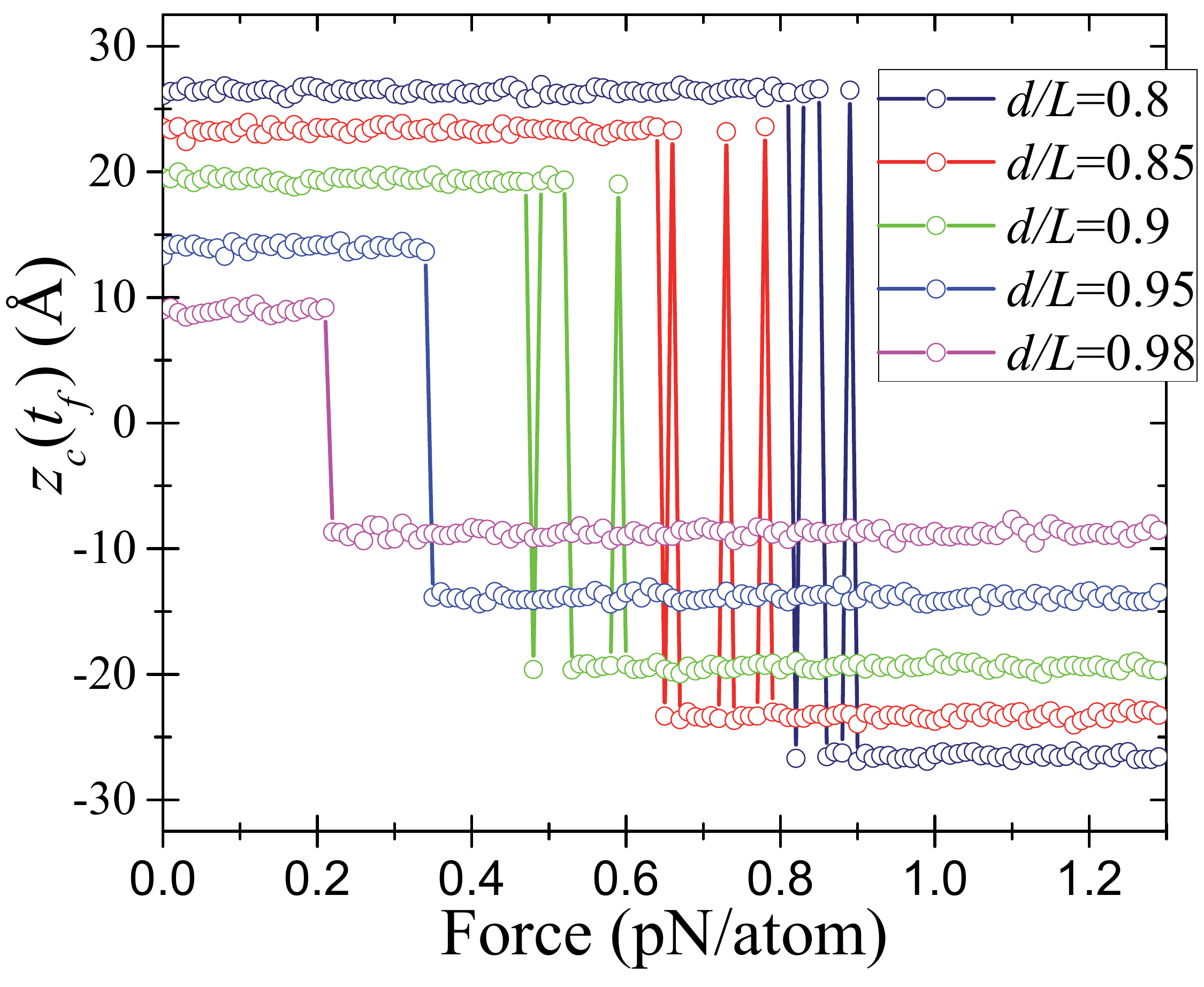}}
\caption{Up-to-down transition: the final position (at a time $t_f$) of a
central atom of membrane as a function of the applied force magnitude for
(a) shorter membrane \textbf{\textit{A}} (22 rings length), and (b) longer
membrane \textbf{\textit{B}} (42 rings length). The calculation details are
given in the text. }
\label{fig4}
\end{figure}

In order to simulate the up-to-down transition, the double-clamped graphene
nanoribbon buckled upwards was subjected to the force in the downward
direction. The final state of membrane was found using MD simulations as
described in SI Sec. ”MD Simulation details”. Figure~\ref{fig4} shows the
final position of a central atom of membrane versus the applied force for
several values of $d/L$ and two membrane sizes. According to Fig.~\ref{fig4},
at fixed $L$, the up-to-down threshold switching force (the minimal force
required for the up-to-down transition) is larger for smaller values of $d/L$.
Moreover, at fixed $d/L$, the threshold switching force is smaller for
longer membranes. Additionally, some curves in Fig. \ref{fig4} exhibit a
noisy threshold (such as $d/L=0.8$ and $d/L=0.98$ in (a)), which can be
related to thermal fluctuations of membranes. All these observations are
intuitively reasonable.

\begin{figure}[tbh]
\centering{(a)\includegraphics[width=0.42\columnwidth]{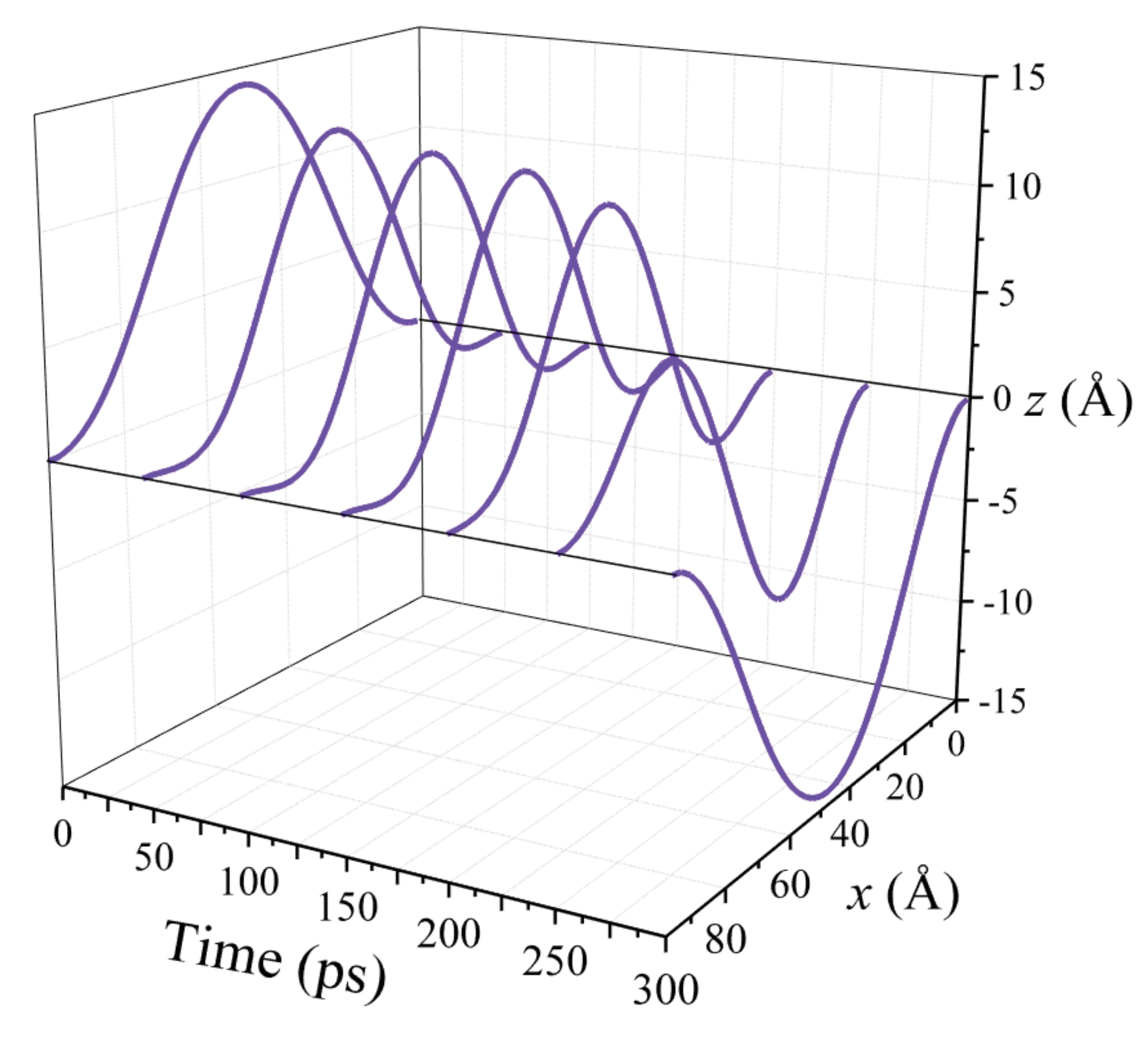} \qquad (b) %
\includegraphics[width=0.42\columnwidth]{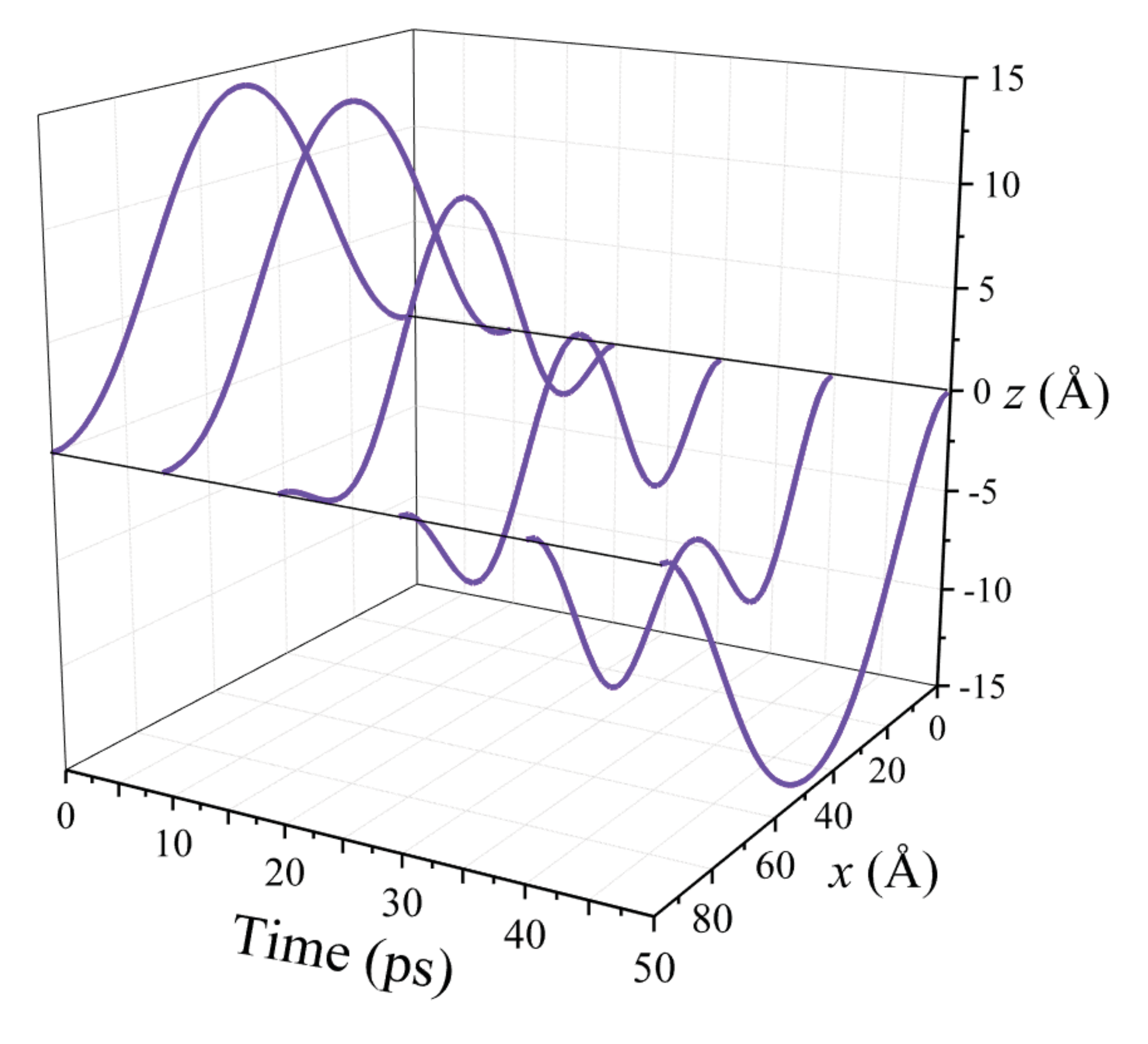}}
\caption{Geometries of membrane \textbf{\textit{B}} in the process of
the up-to-down switching at smaller ($F=0.5$ pN/atom) (a) and larger ($F=1.49$ pN/atom) (b) forces.
These geometries were obtained using overdamped MD simulations for $d/L=0.95$ (the simulation parameters are provided in the Supplementary Information). The switching occurs through the non-symmetric path at
smaller forces (exceeding the threshold) (a), and
symmetric path at larger forces (b).}
\label{fig3}
\end{figure}

Depending on the force magnitude, the membrane switching occurs either
through the symmetric or non-symmetric membrane profile (see Fig. \ref{fig3}).
Figure \ref{fig3} was obtained using overdamped simulations of membrane dynamics
(additional results of these simulations can be found in SI Sec. ”MD Simulation details”).
The non-symmetric profile is associated with a smaller
energy barrier~\cite{Sedelnikova2016} and involved in switching by smaller
forces. Larger applied forces result in the switching through the symmetric
profile. According to our observations and previous work~\cite%
{Sedelnikova2016}, in all cases, the membrane profile is symmetric at short
times. If the force magnitude is sufficient to overcome the energy barrier
for the symmetric profile, then, typically, the switching takes place through the
symmetric path. Otherwise, a symmetry breaking occurs leading to the
switching through the lower energy barrier associated with the non-symmetric
membrane profile. Thermal fluctuations help the symmetry breaking.


\subsection{Down-to-up transition}\label{sec2c}

It was suggested in Ref. \citenum{pershin11c} that the memcapacitor
membrane can be set into the buckled upwards state 0 (see Fig. \ref%
{fig1}) by also using the electrostatic attraction between the capacitor plates. When
the pulled-down membrane is suddenly released, there are conditions such
that the membrane overcomes the potential barrier and ends up in the buckled
upwards state. As the kinetic energy plays an important role in the down-to-up transition, this
process can not be analyzed using the overdamped dynamics.

\begin{figure}[t]
\centering{(a)\includegraphics[width=0.42\columnwidth]{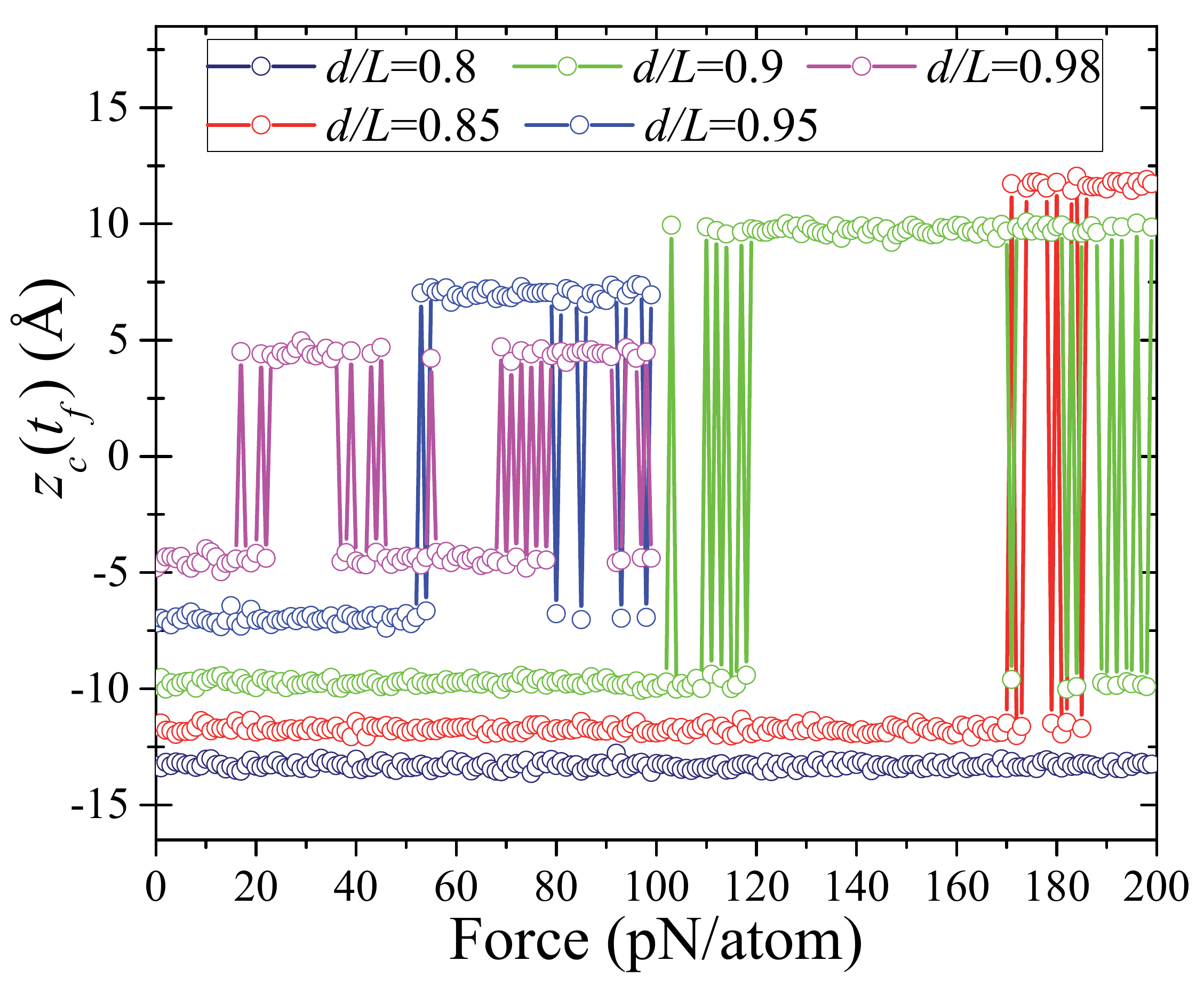} (b)%
\includegraphics[width=0.42\columnwidth]{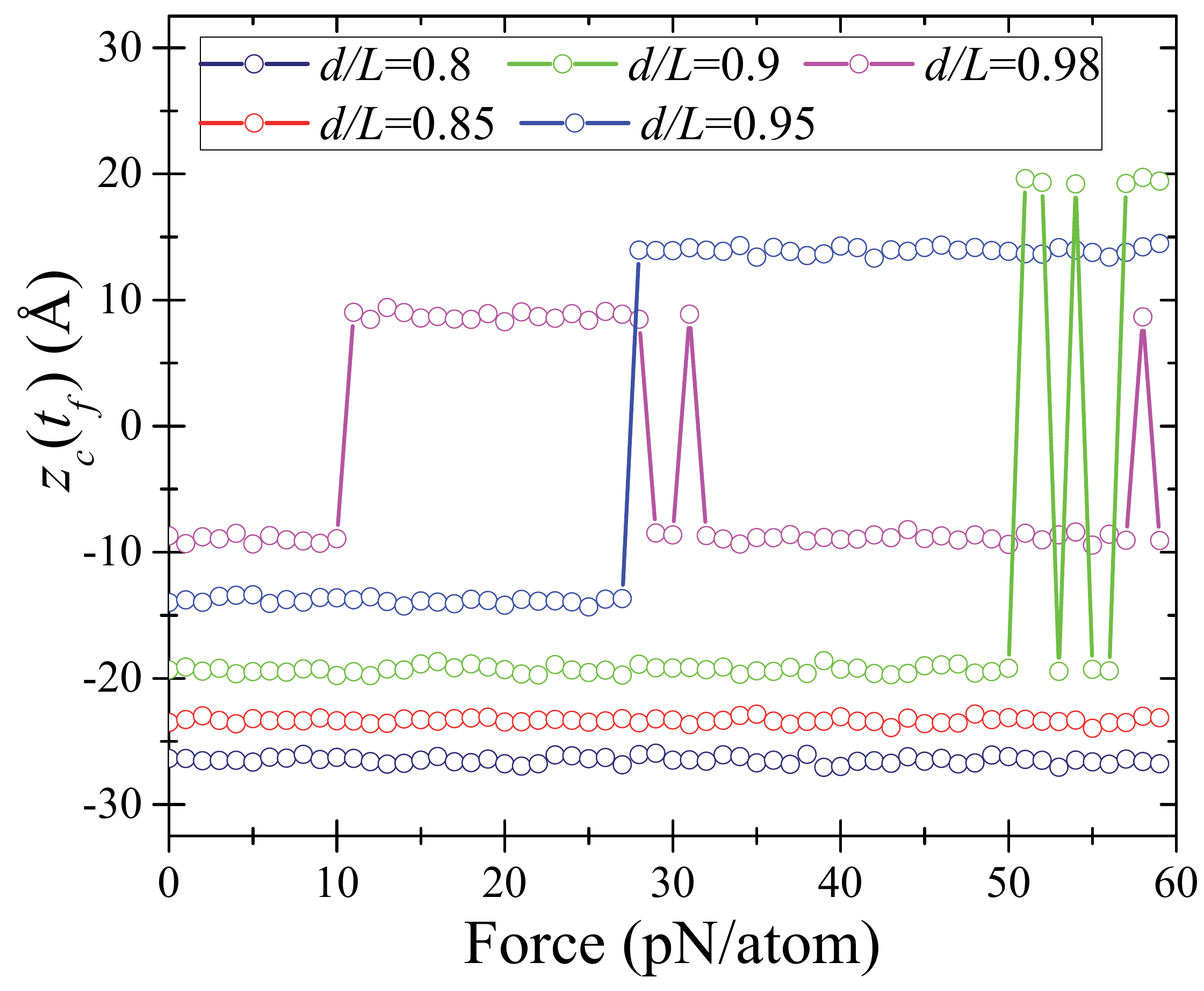}}
\caption{Down-to-up transition: the final position of a central atom of
membrane as a function of the applied force magnitude for (a) shorter
membrane \textbf{\textit{A}} (22 rings length), and (b) longer membrane
\textbf{\textit{B}} (42 rings length).}
\label{fig6}
\end{figure}

In order to simulate the down-to-up transition, every atom of a double-clamped
buckled membrane was subjected to a constant force in -$z$ direction. After
an equilibration period, the forces were removed, and the system was
simulated for a time interval sufficient to reach the steady
state. The final position of a central atom of membrane is presented in Fig. %
\ref{fig6} as a function of the applied force for several values of $d/L$.

Figure~\ref{fig6} shows that the threshold forces for the down-to-up
transition are larger than those for the up-to-down transition (see Fig. \ref%
{fig4}). Moreover, the threshold forces for the down-to-up transition are
larger for the shorter membrane \textbf{\textit{A}} compared to the longer
membrane \textbf{\textit{B}}. It is interesting that the final state of
membrane oscillates as a function of the applied force (see $d/L=0.98$
curves in Fig. \ref{fig6}). Qualitatively, having a high kinetic energy, the
membrane moves up and down while its energy dissipates. Finally, it gets
trapped in one of two stable buckled states.

Overall, our MD results indicate the possibility of writing one bit of
information into the state of membrane memcapacitor. In this
interpretation, logic 1 corresponds to the membrane buckled downwards, and
logic 0 - to the membrane buckled upwards.
In addition to $0\rightarrow 1$ and $1\rightarrow 0$ transitions presented in Figs. \ref{fig4} and \ref{fig6},
we have verified the occurrence
of $1\rightarrow 1$ and $0\rightarrow 0$ transitions at the same values of
forces required for $0\rightarrow 1$ and $1\rightarrow 0$ transitions,
respectively. Therefore, logic 0 or 1 can be written into the memcapacitor
just by selecting a suitable value of the applied force.

In our work, we considered relatively small nanoribbons, which are mechanically rigid. This is an important requirement for the memcapacitor application as the mechanical nonvolatile information storage is not possible with flexible graphene. From some previous studies~\cite{Schoelz15} it is known that local structural corrugations (ripples)~\cite{Fasolino07a} disappear in the transition from flexible to rigid graphene. In agreement with this previous work~\cite{Schoelz15}, our molecular dynamics simulations have shown the absence of ripples (spontaneous height fluctuations) in nanoribbons of reported sizes and their existence in larger nanoribbons. The latter, however, are not suitable for the memcapacitor application because of their flexibility. To summarize, while our simulation tools provide a means for ripples detection, we did not observe these in the reported structures that are mechanically rigid.

\section{Theory of Elasticity}

Even though the graphene has a thickness of one atom, the graphene membranes
are quantitatively good described by the theory of elasticity \cite{Eriksson13, Tomi15, Abdi16, Bonilla16}. This allows us, on one hand, to
obtain analytical expressions for the buckled membranes~\cite{Qiu04,
Cazottes09, Bitafaran15} and, on the other hand, perform a comparison with
 MD simulations.

There is a number of publications dealing with buckled beams and membranes
under the transverse load (see, for example, Refs.~[\citenum{Medina2014,Chen2015,Krylov2008,Medina2012}]).
Such systems are frequently described in the framework of Euler-Bernoulli
theory, which, however, leads to complex analytically unsolvable equations.
Bubnov-Galerkin decomposition is a one of the best approaches to find the
approximate analytical solution of these equations, although it still
requires a complex phase-diagram analysis. In particular, using such
analysis, the authors of Refs.~[\citenum{Medina2014,Chen2015,Krylov2008,Medina2012}] investigated an
electrostatically loaded double-clamped membrane above a rigid flat
electrode and derived cumbersome conditions for the symmetric snap-through
transition, symmetry breaking, existence of bifurcation and pull-in
instability.

In this Section, we derive compact but sufficiently precise expressions for
the snap-through switching forces based on the theory of elasticity.

\subsection{Buckling and snapping-through within~the~theory~of~elasticity}\label{Sec:Elasticity}

\subsubsection{Description of buckled membranes within the theory of elasticity}

Consider a 2D membrane within the theory of elasticity \cite{Abdi16, Tomi15,
LL}. The total potential energy of membrane is defined by the deflection
$\zeta $ (along the normal direction $z$) as
\begin{equation}
U=\frac{D}{2}\iint \textnormal{d}S\left( \Delta \zeta \right) ^{2}+T\delta S+\iint
\textnormal{d}SF\zeta .  \label{U}
\end{equation}%
Here, $\Delta $ stands for the 2D Laplacian, $T=T_{0}+\delta T$ is the total
tensile force, $T_{0}$ is the force applied by the support and $\delta T$ is
the bending tension resulted from the extension,%
\begin{equation}
\delta T=E_{2\mathrm{D}}\frac{\delta S}{S_{0}},\qquad \delta S=\iint
\textnormal{d}S\left( \nabla \zeta \right) ^{2},  \label{DT}
\end{equation}%
$S_{0}=wL$, and $F$ is the external force density. The compression of membrane
corresponds to $T_{0}<0$.

Given the potential energy, one may also be interested in the dynamics of
membrane, which is defined by the equation
\begin{equation}
\mu \frac{\partial ^{2}\zeta }{\partial t^{2}}+\mu \gamma \frac{\partial
\zeta }{\partial t}-F=-D\Delta \Delta \zeta +T\Delta \zeta ,  \label{eq}
\end{equation}%
where $\mu $\ is the 2D mass density.

\subsubsection{Eigenmodes and buckling}

The spatial harmonics (eigenmodes) of clamped membranes can be written
as
\begin{eqnarray}
\zeta _{n}(x) &=&\cosh \left( b_{n}\frac{x}{d}\right) -\cos \left( b_{n}\frac{x}{d}\right) -  \label{zeta_n}\frac{\cosh b_{n}-\cos b_{n}}{\sinh b_{n}-\sin b_{n}}\left( \sinh \left(
b_{n}\frac{x}{d}\right) -\sin \left( b_{n}\frac{x}{d}\right) \right) ,
\end{eqnarray}%
where the numbers $b_{n}$ are the solutions of the equation $\cosh b_{n}\cos
b_{n}=1$. One can find that $b_{n}\approx 3\pi /2+n\pi $ (in particular, $%
b_{0}=4.73$ is close to $3\pi /2=4.65$). The functions $\zeta _{n}(x)$ are
orthonormal.

Consider a buckled membrane shown in Fig. \ref{fig1}.
Its potential energy (Eq. (\ref{U})) calculated for the fundamental $n=0$ mode (taking $\zeta (x)=\zeta _{0}(x)$
with $\zeta (d/2)\equiv \zeta _{\mathrm{c}}$) is
\begin{equation}
U=-\alpha \zeta _{\mathrm{c}}^{2}+\beta \zeta _{\mathrm{c}}^{4}+f\zeta _{%
\mathrm{c}},  \label{U_buckled}
\end{equation}%
where $\alpha = \pi ^{2}w \left( \left\vert
T_{0}\right\vert -T_{\mathrm{c}}\right) /\left(4d\right)$, $\beta =3\pi ^{4}
Dw/(4\varepsilon ^{2}d^{3})$, and $f=\pi Fdw/6$. Here, $T_{\mathrm{c}}=4\pi ^{2}D/d^{2}$
is the critical tension. At $\alpha >0$, that is at $%
\left\vert T_{0}\right\vert >T_{\mathrm{c}}$, the potential is the
double-well potential with minima at symmetric deflections, $\zeta _{c}=\pm
\sqrt{\alpha /2\beta }$, and the potential barrier $\delta U=\alpha
^{2}/4\beta $. (We note that a less realistic case of hinged boundary
conditions would lead to simpler expressions for eigenmodes, $\sim \sin \left( \pi nx/d\right)$,
compared to Eq.~(\ref{zeta_n}), making the hinged boundary conditions preferable for some calculations.
However, this case results in quantitatively different values. For example,
the critical tension differs four times.)

We note that the nanoribbon length is given by
\begin{equation}
L=\int_{0}^{d}\mathnormal{d}x\sqrt{1+\zeta ^{\prime 2}}.
\end{equation}%
Expanding this expression to the second order, one obtains the interrelation
between $L$, $d$, and the maximal deflection $\zeta _{\mathrm{c}}$:%
\begin{equation}
\zeta _{\mathrm{c}}\approx 0.64\sqrt[8]{\frac{d}{L}}\sqrt{L(L-d)}.
\label{z0}
\end{equation}%
On the other hand, the maximal deflection of buckled membrane can be
inferred from the minima of the potential energy, Eq.~(\ref{U_buckled}):
\begin{equation}
\zeta _{\mathrm{c}}^{2}=\frac{\alpha }{2\beta }=\frac{2}{\pi ^{2}}\frac{L^{2}%
}{E_{2\mathrm{D}}}\left( \left\vert T_{0}\right\vert -T_{\mathrm{c}}\right) .
\label{zc}
\end{equation}%
Equations~(\ref{z0}) and (\ref{zc}) link the tensile force $T_{0}$ to the
parameter $d/L$, which  can be considered as a measure of deformation of buckled membrane.
These equations can be used to express the tensile force $T_{0}$ through $d/L$.

\subsubsection{Snap-through transition}

Here we present a convenient method to describe the dynamics of membrane subjected to an external force and find the
minimal force causing its snap-through transition. The main idea is to expand $\zeta (x,t)$ in
harmonics $\zeta _{n}(x)$\ (given by Eq. (\ref{zeta_n})) with amplitudes $q_{n}(t)$ as
\begin{equation}
\zeta (x,t)=\sum_{n}q_{n}(t)\zeta _{n}(x),  \label{expansion}
\end{equation}%
and limit the sum to few first terms. We found that in order
to describe the symmetric and non-symmetric transitions, it is sufficient to
consider $n=0,2$ and $n=0,1$ terms in Eq. (\ref{expansion}), respectively. The calculation consists
in the following. First, we substitute the expansion (\ref{expansion}) in Eq.~(\ref%
{eq}). Second, the resulting equation is multiplied by $\zeta _{m}(x)$ and integrated
taking into account the orthogonality of harmonics. Here we emphasize that for
the harmonics (\ref{zeta_n}), the integrals are readily calculated,
and instead of the integro-differential equation one obtains a system of
differential equations for the functions $q_{n}(t)$. Inserting the obtained
functions $q_{n}(t)$ back into Eq.~(\ref{expansion}) gives us the description
of membrane dynamics. In the following, the results of these
calculations (performed using the time normalized by the
membrane characteristic frequency $\omega _{%
\mathrm{c}}=(1/L^{2})\sqrt{D/ \mu }$ and $t_{\mathrm{c}%
}=1/\omega _{\mathrm{c}}$)~\cite{Gomez17} are analyzed for membrane \textbf{\textit{B}}.

\begin{figure}[t]
\centering{\includegraphics[width=0.7\columnwidth]{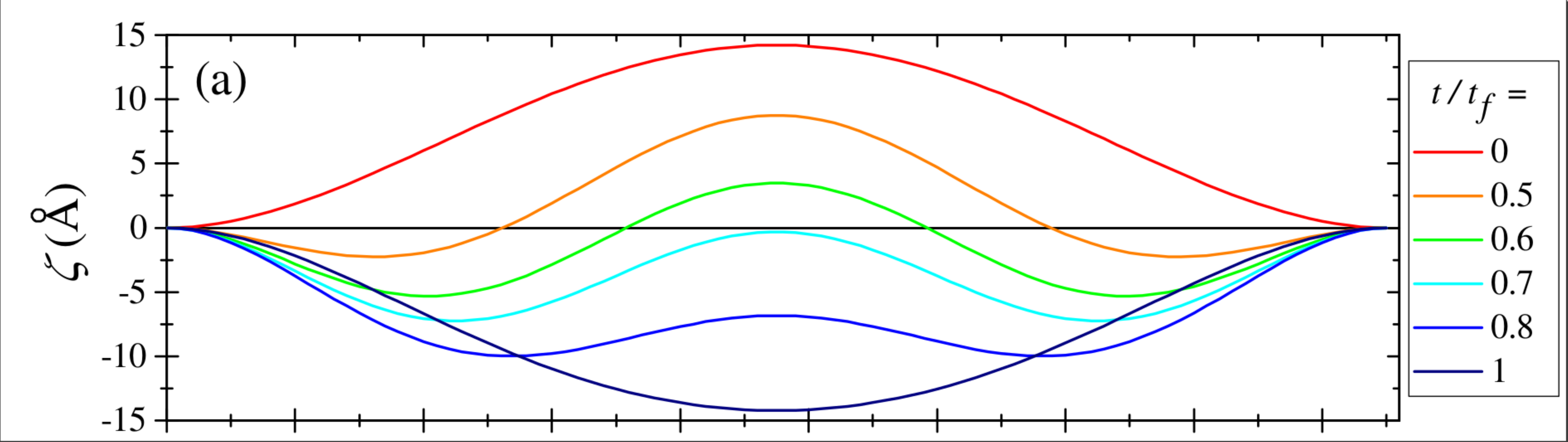} }\includegraphics[width=0.7	%
\columnwidth]{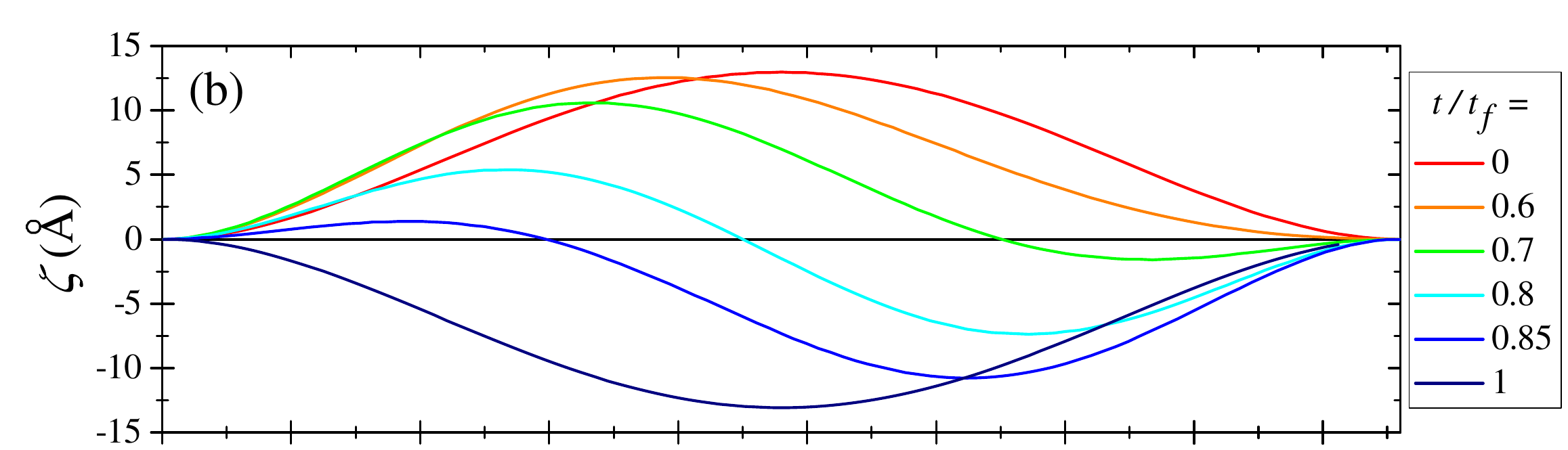} \includegraphics[width=0.7\columnwidth]{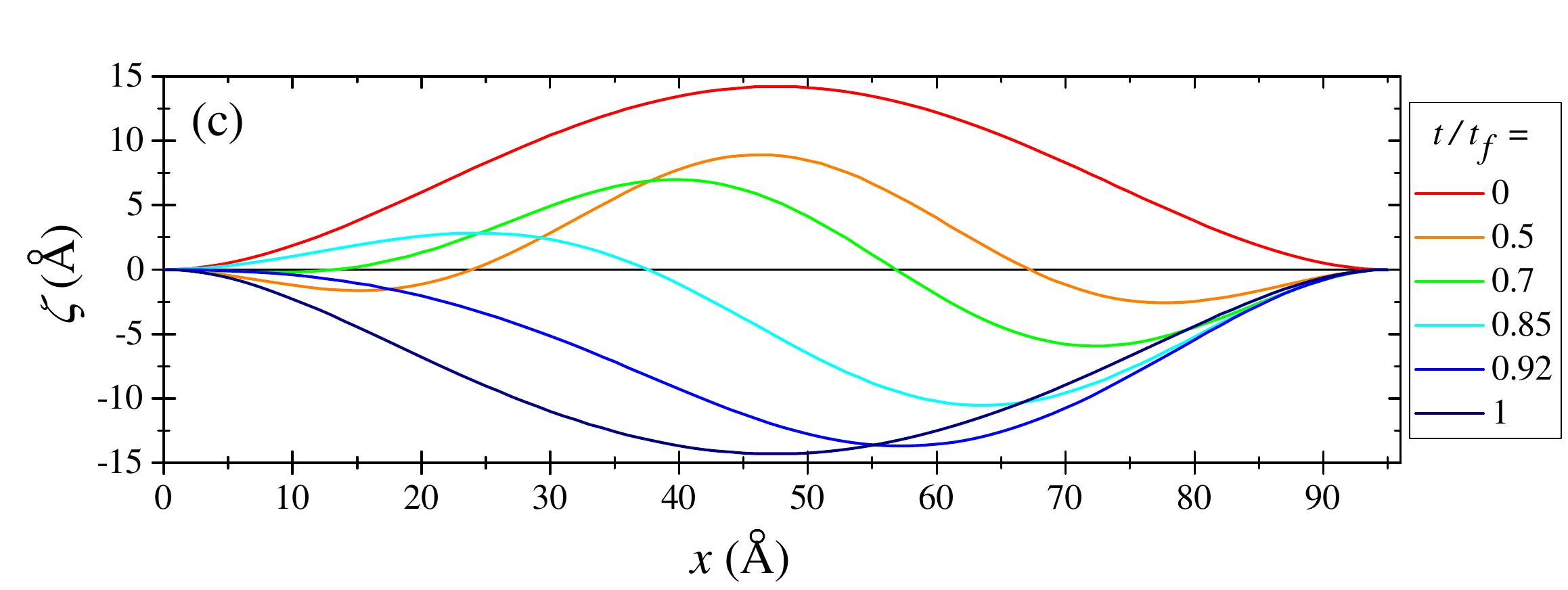}
\caption{Up-to-down transition within the theory of elasticity.
(a)~Fully symmetric switching. (b) Non-symmetric
switching caused by a small asymmetry in the applied force. (c) More
realistic switching scenario based on a symmetric force, where the initial
symmetric distortion becomes non-symmetric at longer times. For all panels $%
d/L=0.95$.}
\label{Fig:i}
\end{figure}


Consider the results presented in  Fig.~\ref{Fig:i}.
Fig.~\ref{Fig:i}(a) demonstrates the case of fully symmetric switching, when the dynamics can be described by
$n=0$ and $2$ harmonics. Next, we introduce a small asymmetry into the
problem via a small non-symmetric modification of the force (of the order of $0.1\%$). This results in the non-symmetric dynamics of
Fig.~\ref{Fig:i}(b), which can be described in terms of $ n=0$ and $1$ harmonics.
 We note that the force needed for the non-symmetric
snap-through transition is smaller than that for the symmetric one,
which will also be analyzed below in more detail. Finally, in Fig.~\ref{Fig:i}(c) we illustrate
a combination of the above regimes found using a symmetric force, when
a tiny fluctuation in the numerical solution changes the symmetric dynamics to the
non-symmetric one. Note that this case is close to the one discussed in Ref.~\citenum{Sedelnikova2016}.

\begin{figure}[t]
\centering{\includegraphics[width=0.49\columnwidth]{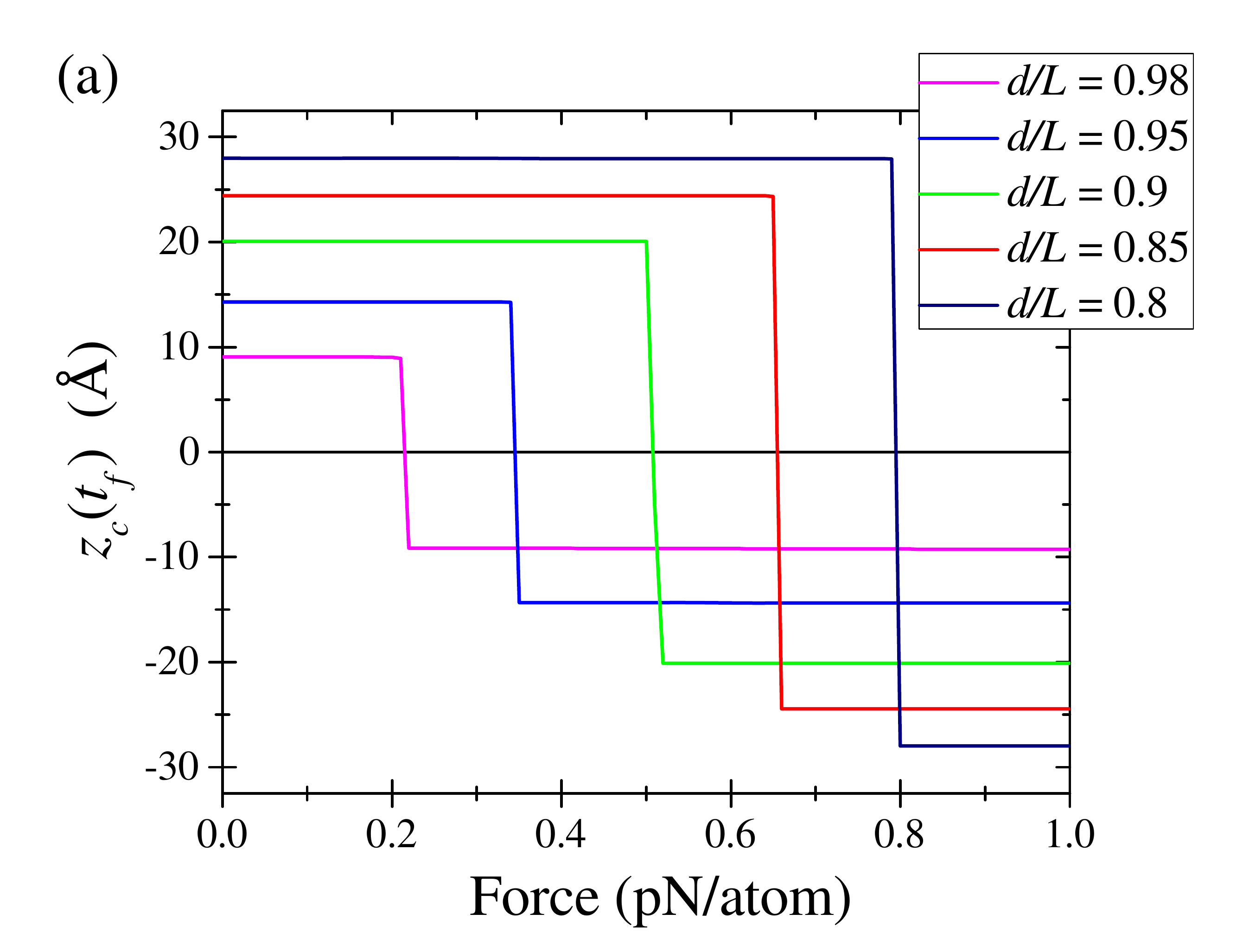}
\includegraphics[width=0.49\columnwidth]{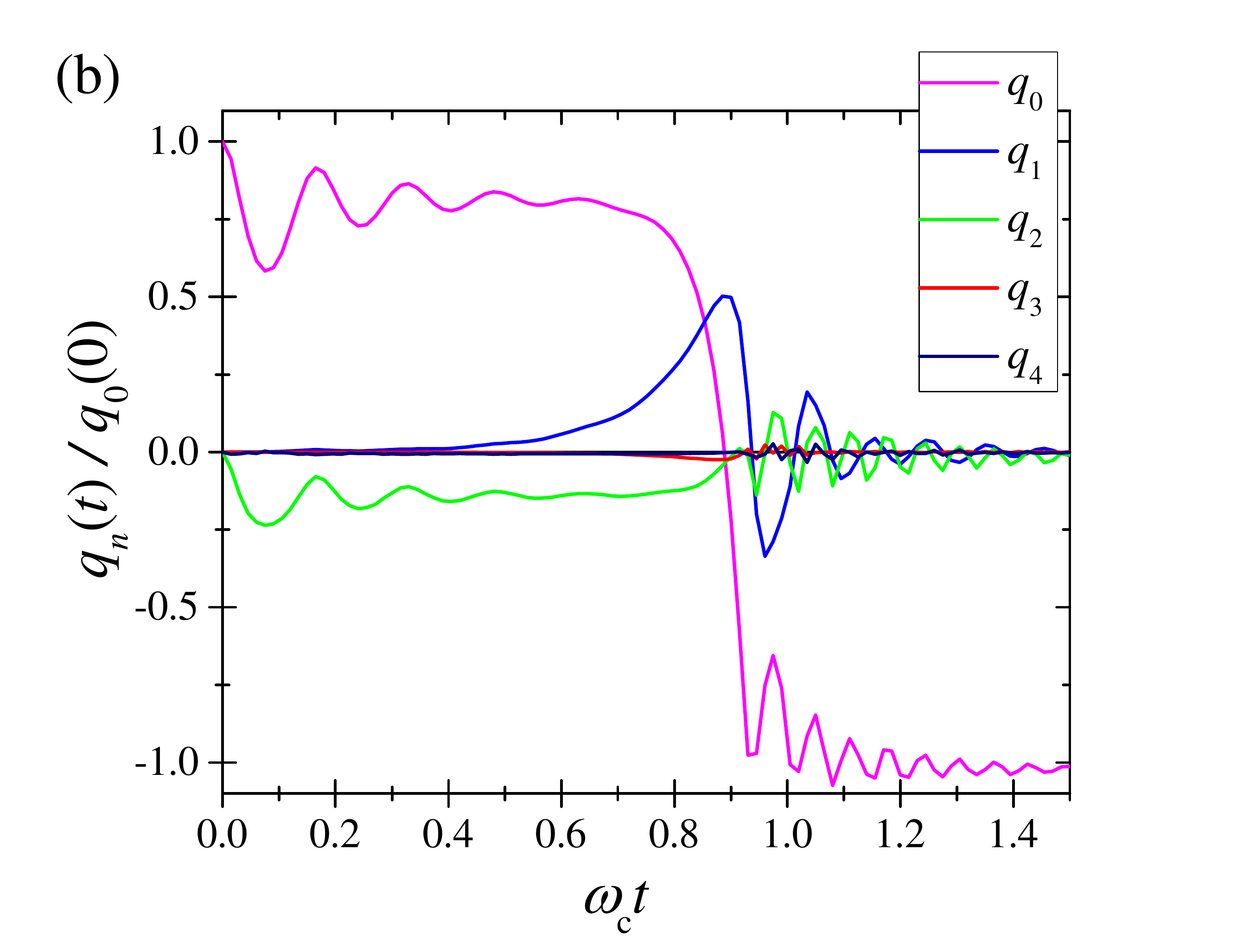}}
\caption{Numerical simulation of up-to-down transition. (a) The final
central-point deflection $\protect\zeta _{\mathrm{c}}$ versus the applied
transverse force $F$ for several values of $d/L$. (b) Time-dependence of the
first few harmonics' amplitudes $q_{n}(t)$.}
\label{Fig:ii}
\end{figure}

Figure~\ref{Fig:ii}(a) depicts the dependence of the final position of
membrane center, $\zeta _{\mathrm{c}}(t_{f})$, on the
applied force $F$. At a certain value of force, $F=F^{\downarrow }$, the
membrane switches from the buckled upwards state to the buckled downwards state. Our
calculations show that the threshold force is proportional to the initial
central-point deflection $\zeta _{\mathrm{c}}$ and is different for the symmetric (s) and non-symmetric (ns) dynamics. Namely, this force, being multiplied by the membrane area $wL$, reads
\begin{eqnarray}
\frac{F_{\mathrm{s,ns}}^{\downarrow }}{F_{0}} &=&C_{\mathrm{s,ns}}\sqrt{2.44}%
\frac{\zeta _{\mathrm{c}}}{d}=C_{\mathrm{s,ns}}\left( \frac{L}{d}\right)
^{3/8}\sqrt{\frac{L}{d}-1},  \label{F_star} \\
F_{0} &=&\frac{Dw}{L^{2}},\qquad C_{\mathrm{ns}}=253.4,\qquad C_{\mathrm{s}%
}=359.1.  \label{Asa}
\end{eqnarray}%
Since the non-symmetric transition is energetically favorable, we expect that this value of force is the one to be
used in the device design/experiment analysis. We note that estimations based on Eq. (\ref{F_star}) are in a
good agreement with results found by other methods as we discuss
later and illustrate in Fig.~\ref{fig9}. Figure~\ref{Fig:ii}(b) presents the time-dependence of harmonics' amplitudes $q_{n}(t)$
in the up-to-down transition. Importantly,
the dynamics is well described by  $n=0$, $1$, and $2$ harmonics, while $%
n=3$ and $4$ harmonics amplitudes are negligibly small.

Our calculations (similar to the consideration of the up-to-down force above) demonstrate that the minimal force needed for the down-to-up
transition is proportional to $\zeta _{\mathrm{c}}^{2}$ as
\begin{eqnarray}
\frac{F^{\uparrow }}{F_{0}} &=&C^{\uparrow }2.44\left( \frac{\zeta _{\mathrm{%
c}}}{d}\right) ^{2}=C^{\uparrow }\left( \frac{L}{d}\right) ^{3/4}\left(
\frac{L}{d}-1\right) ,  \label{Fup} \\
C^{\uparrow } &=&8.48\cdot 10^{4}.
\end{eqnarray}%
See also SI Sec. ”Down-to-up transition”, where the down-to-up transition is
illustrated graphically.

\subsection{Phenomenological elasticity theory}

\label{sec3}

In this Section, in order to avoid all complications associated with the
Bubnov-Galerkin decomposition, we elaborate a different approach describing
the up-to-down and down-to-up transitions in a compact analytical form.

\subsubsection{Ansatz}

\label{sec3a}

Following our observation of the existence of the symmetric and
non-symmetric membrane profiles in the snap-through transition (see Fig.~\ref%
{fig3}), we introduce two simplest polynomial functions to describe such
symmetric (s) and non-symmetric (ns) shapes of membrane:
\begin{eqnarray}
\theta _{\text{s}}(s) &=&A_{\text{s}}\left( s^{2}-\frac{1}{4}\right) s\left(
s^{2}-c_{0}^{2}\right) ,  \label{eq:theta_s} \\
\theta _{\text{ns}}(s) &=&A_{\text{ns}}\left( s^{2}-\frac{1}{4}\right) \left(
s-c_{1}\right) \left( s-c_{2}\right) .  \label{eq:theta_ns}
\end{eqnarray}%
Here $\theta _{i}(s)$ is the angle that the membrane makes with the horizontal, $%
s$ is the internal coordinate that changes between $-1/2$ and $1/2$, $A_{i}$
and $c_{i}$ are coefficients, and $i=\{\text{s},\text{ns}\}$. Clearly, Eqs. (\ref%
{eq:theta_s}) and (\ref{eq:theta_ns}) describe the double-clamped membrane
as $\theta _{i}(\pm 1/2)=0$.

\subsubsection{Up-to-down transition}

\label{sec3b}

Our goal here is to estimate the minimal force required for the snap-through
transition. In this subsection, we assume that the snap-through transition
is induced by a slowly increasing force such that the system always stays in
the potential energy minimum. At zero applied force, there are two minima of
the potential energy corresponding to the two stable states of membrane. The
applied force modifies the potential energy landscape such that, starting at
a certain force, the potential energy has a single minimum. Here, this value
of force is found and considered as an estimation for the threshold
switching force.

In the presence of a constant force $F$ in the downward direction, the
membrane potential energy is given by
\begin{equation}
U=U_{b}+U_{str}+U_{ext}=\frac{Dw}{2L}\int\limits_{-1/2}^{1/2}\left[ \frac{%
\partial \theta} {\partial s} \right]^2\textnormal{d}s+U_{str}+U_{ext},
\label{eq:pot_energ}
\end{equation}
where $U_b$ is the bending energy, $U_{str}$ is the stretching energy, $%
U_{ext}$  is the potential energy due to the applied force. Then, using $U_{ext}=F z_{cm}$, where $z_{cm}$ is $z$
coordinate of the center of mass, and neglecting $U_{str}$ term not
important for the up-to-down transition (see, e.g., Fig.~S1 that
indicates the insignificance of $U_{str}$ in buckled membrane), we get the
following equation for the potential energy extrema:

\begin{equation}
F=-\frac{\text{d} U_b}{\text{d} z_{cm}}=-\frac{\text{d} U_{b}}{\text{d}c_i}%
\left(\frac{\text{d}z_{cm}}{\text{d}c_i}\right)^{-1},  \label{eq22}
\end{equation}
where $i=\{0,1\}$ corresponds to  $\{\text{s},\text{ns}\}$,  respectively. Equation (\ref{eq22}) allows
finding $c_0$ and $c_1$ for a given strength of the applied
force.

The minimal force needed for the snap-through transition corresponds to the
maximum of $F(z_{cm})$ and, for the transition through the symmetric shape,
is given by
\begin{equation}
F^\downarrow_\text{s}=440\frac{Dw}{L^2}\sqrt{\frac{L-d}{L}} \label{eq20}
\end{equation}
taking place at $c_0=0.3589$ (the corresponding membrane profile is
presented in Fig.~S4). At this value of $c_0$, the bending energy is
$U_{b,\text{s}}=108Dw(L-d)/L^2$.

Performing the same calculations for the non-symmetric shape, one can find
that the force needed to support the equilibrium non-symmetric shape
decreases with the progress of switching. A zero force is reached at $%
c_{1}=-c_{2}=1/(2\sqrt{5})$ that corresponds to the maximum of $%
U_{b,\text{ns}}=90Dw(L-d)/L^{2}$. A possible (rough) estimation for the threshold
switching force can be obtained taking the limit $c_{1}\rightarrow \infty $
leading to
\begin{equation}
F_{\text{ns}}^{\downarrow}=281\frac{Dw}{L^{2}}\sqrt{\frac{L-d}{L}}  \label{eq888}
\end{equation}%
and $U_{b,\text{ns}}=42Dw(L-d)/L^{2}$.

In fact, a realistic scenario of the switching through the non-symmetric
shape can be imaged as follows. We start with a symmetric membrane at $F=0$
and slowly increase the force. The symmetry breaking occurs at a certain
value of force, say, $F^{\downarrow}$. The dynamics of switching is a complex
process significantly relying on thermal fluctuations. As the switching
dynamics can not be reached within the framework of our model, for
estimation purposes, we assume that $F^{\downarrow}$ corresponds to the point where
the bending energies and applied forces for the symmetric and non-symmetric
shapes are the same. One can find that both conditions are satisfied at $%
c_0=0.4683$ and $c_1=0.6948$, so that $U_b=47.8Dw(L-d)/L^2$ and
\begin{equation}  \label{eq:fsw_nonsym}
F^\downarrow=263.53\frac{Dw}{L^2}\sqrt{\frac{L-d}{L}}.
\end{equation}
Consequently, at $F<F^{\downarrow}$, the membrane keeps the symmetric shape and
switches to the non-symmetric one as soon as $F$ reaches $F^{\downarrow}$. As there
is no barrier involved in the non-symmetric switching, no further increase
in the applied force is required to complete the snap-through transition
through the non-symmetric shape.

\subsubsection{Down-to-up transition}

\label{sec3d}

As the stretching energy $U_{str}$ (see Eq. (\ref{eq:pot_energ})) plays an important role in the down-to-up transition, it needs to be taken into account. For our purposes, it is sufficient to approximate $U_{str}$ as
\begin{equation}
U_{str} = \frac{E_{2D} w}{2 L} \Delta L^2 ,  \label{eq437}
\end{equation}
where $\Delta L$ is the change in the membrane length. In the approximation of small elongation, $\Delta L\ll L$, one can assume that both $U_b$ and the shape of the stretched membrane are not significantly modified compared to $F=0$ case. In this situation,
at the threshold of transition, the stretching energy (Eq. (\ref{eq437})) is equal to the difference between the maximal and minimal bending energies (the energy conservation condition).
Using the energies given below Eq. (S8) we find
\begin{equation}  \label{eq:dtu_energ}
\frac{E_{2D} w}{2 L} \Delta L^2=U_b^{max}-U_b^{min} = 161 D w \frac{L-d}{L^2}.
\end{equation}

Moreover, at equilibrium
\begin{equation}
F(z_{cm})=\frac{E_{2D}w}{2L}\cdot \frac{\text{d}(\Delta L^{2})}{\text{d}
z_{cm}}.  \label{eq:dtu_force}
\end{equation}%
Under the condition of small $\Delta L\ll L$, the center of mass position of
membrane (buckled downwards) can be expressed as
\begin{equation}
{z_{cm}}^{\ast }=-0.3187\sqrt{L(L-d)}+\frac{\text{d}z_{cm}}{\text{d}(\Delta
L)}\Delta L,  \label{eq:z_cm1} \\
\end{equation}%
with
\begin{equation}
\frac{\text{d}z_{cm}}{\text{d}(\Delta L)}=0.3187\frac{2L-d}{2\sqrt{L(L-d)}}.
\label{eq:z_cm2}
\end{equation}%
Using Eqs. (\ref{eq:dtu_energ}) and (\ref{eq:z_cm2}), Eq. (\ref{eq:dtu_force}%
) can be rewritten as
\begin{equation}
F^\uparrow=79.62\sqrt{2DE_{2D}}\frac{w}{L}\frac{L-d}{2L-d}.  \label{FZCM}
\end{equation}%
We have verified that the threshold switching force for the down-to-up
transition given by Eq. (\ref{FZCM}) is in agreement with our typical MD
simulations results. A very good agreement was obtained with MD simulations
performed with very small energy dissipations.

\section{Density Functional Theory}\label{Sec:DFTmain}

A DFT investigation of the up-to-down transition was
performed following the previously reported study of the
up-to-down switching of buckled graphene membrane~\cite{Sedelnikova2016}.
In the present calculations, we focused on the effects of boundary conditions and
buckling strength on the up-to-down threshold switching force.
DFT computational details can be found in SI Sec. "DFT Calculations".

\begin{figure}[t]
	\centering{\includegraphics[width=0.40\columnwidth]{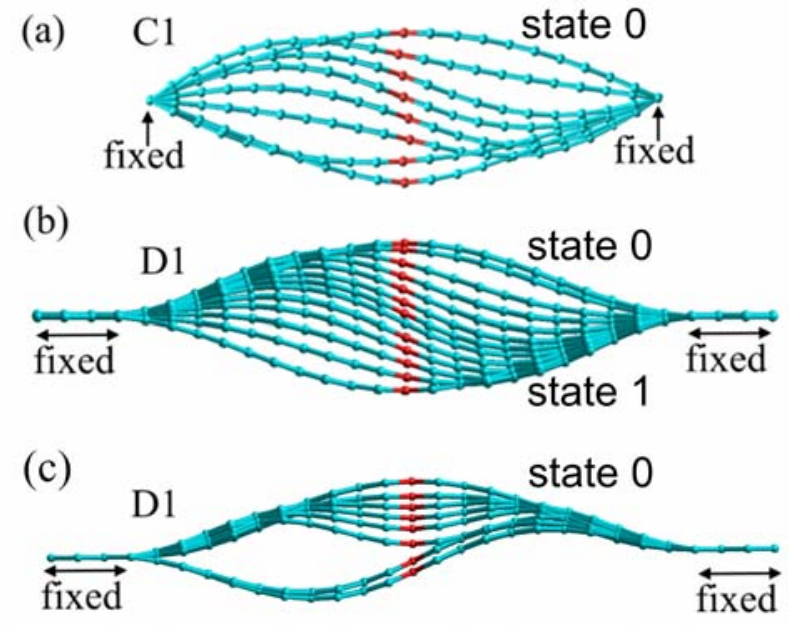}}
	\caption{Geometries of membranes C1 (hinged boundary conditions, (a)) and D1 (clamped boundary, (b) and (c))
		in the process of the up-to-down switching. The switching occurs through non-symmetric paths (a) and (b), and initially symmetric path (c). These geometries were found using DFT optimization of geometries obtained utilizing a progressive displacement of central atoms.}
	\label{Fig:DFT1}
\end{figure}

We investigated a buckled zigzag graphene nanoribbon specified in SI.
Two types of boundary conditions
were examined: 1) hinged boundary conditions (configurations C1 with $d/L=0.96$, Fig.~\ref{Fig:DFT1}(a),
C2 with $d/L=0.78$, and C3 with $d/L=0.63$), and 2) clamped boundary
conditions (configuration D1 with $d/L=0.96$, Fig.~\ref{Fig:DFT1}(b) and (c)).
The initial geometries of nanoribbons corresponded to the buckled upwards state. In order to simulate the
up-to-down transition, we used the approach developed in  Ref.~[\citenum{Sedelnikova2016}].
Specifically, we performed a series of DFT calculations with the central chain of membrane atoms
(marked red in Fig.~\ref{Fig:DFT1}) gradually displaced in $-z$ direction from its position in the buckled upwards state.
The membrane geometry found at the preceding deformation step was used to build its
subsequent modification. In contrast to Ref.~[\citenum{Sedelnikova2016}], in the present calculations only $z$ coordinates of the central chain of atoms were constrained.

The deformation energies of buckled graphene are reported in SI. Our force estimation shows that the threshold switching force for the up-to-down
transition is about $3.8$ pN/atom for C1, $11.1$ pN/atom for C2, and $16.4$ pN/atom for C3.
Moreover, the non-symmetric switching force for D1 is $3.7$ pN/atom and the symmetric one is about $11.5$ pN/atom.
As the non-symmetric switching forces for D1 and C1 are close to each other, we  use the results for C1-C3 for the order-of-magnitude comparison with MD and elasticity theory results.

\section{Discussion}

\label{Sec:Discussion}

\subsection{Four methods to describe the snap-through transitions}
\begin{figure}[tb]
\centering{\includegraphics[width=0.50\columnwidth]{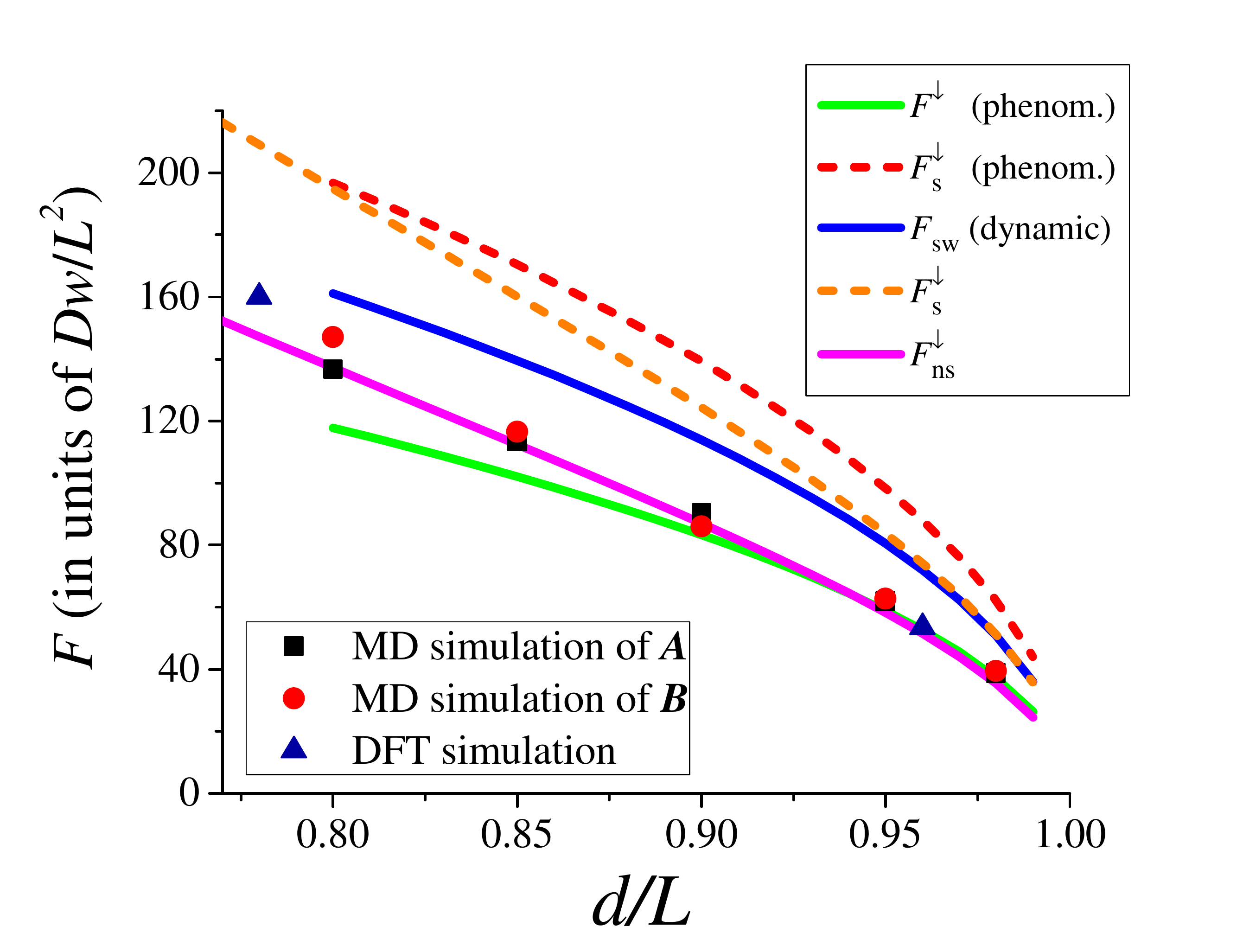}}
\caption{Comparison between the up-to-down threshold forces, calculated by
four different approaches. Green, red, and blue lines \ were calculated in
the non-symmetric, adiabatic and dynamic symmetric switching regimes,
respectively. The results of our MD simulations are shown with squares for
nanoribbon \textbf{\textit{A}} and circles for \textbf{\textit{B}}. The
forces calculated by means of the elasticity theory, Eq.~(\protect\ref%
{F_star}), are shown with the magenta solid line for the non-symmetric
snap-through and with the orange dashed line for the symmetric transition.
The DFT results are presented by triangles.}
\label{fig9}
\end{figure}

As the key finding, Figure~\ref{fig9} presents the dependence of the
up-to-down threshold switching force (calculated with different approaches) on the
membrane deformation parameter $d/L$. In particular, this figure demonstrates
that the threshold forces obtained by a number of different methods are in
good agreement with each other.

We emphasize that:

(i) MD is widely used approach to describe the dynamics of nanoscale
systems. In our work, MD clearly shows the possibility of $0\rightarrow 1$
and $1\rightarrow 0$ transitions of graphene membrane. We have also verified
the occurrence of $1\rightarrow 1$ and $0\rightarrow 0$ transitions at the
same values of forces required for $0\rightarrow 1$ and $1\rightarrow 0$
transitions, respectively.

(ii) In the framework of elasticity theory, the membrane is described
by means of an integro-differential equation for the deflection $\zeta (x,t)$. Expanding the function $\zeta (x,t)$ into  membrane's harmonics allows reducing
the integro-differential equation to a system of differential equations.
This approach was utilized in Sec.~\nameref{Sec:Elasticity}. There it was shown that the
up-to-down threshold switching force depends linearly on the initial central-point deflection, $%
F^{\downarrow }\propto \zeta _{\mathrm{c}}$,\ while the down-to-up
snap-through force at low dissipation depends quadratically on the initial
deflection, $F^{\uparrow }\propto \zeta _{\mathrm{c}}^{2}$. Besides,
we have demonstrated that with a high accuracy, the relevant membrane's dynamics
can be described (and visualized) just by two harmonics.

(iii) The phenomenological approach based on the theory of elasticity has allowed
us to obtain analytical expressions for the up-to-down and down-to-up
switching forces for buckled graphene membrane. The forces for $0\rightarrow
1$ and $1\rightarrow 0$ transitions are in a good agreement with MD
simulations results.

(iv) DFT is a non-standard method for studying mechanical properties of
membranes. Similarly to other methods, DFT  has shown that the non-symmetric
switching pathway is the preferable one. This method also provides a consistent estimation
for the up-to-down threshold switching force.

\subsection{Implications for memcapacitor design}

\label{sec4}

From the application point of view, the membrane memcapacitor should have a
strong 'memory content', namely, the device characteristics in its two logic
states should be sufficiently different so as to provide a significant
influence on other elements of electronic circuits~\cite{diventra13a}. For
this purpose, it is desirable that both the distance between the fixed edge of membrane and
second electrode ($h$ in Fig.~\ref{fig1}) and the maximal deflection of
membrane ($z_{s}(0)$) are chosen (much) smaller than the membrane length.
Moreover, the gap between electrodes should be larger than the maximal
membrane deflection, $h>z_{s}(0)$ (see Eq. (S2) for the definition
of $z_{i}(s)$). For given capacitances $C_{0}$ and $C_{1}$ of the states 0
and 1 (logic 1 corresponds to the membrane buckled downwards, and logic 0 -
to the membrane buckled upwards) such that $C_{1}/C_{0}<3.01$, one can find
that
\begin{equation}  \label{eq27}
h=0.3210\sqrt{L(L-d)}\frac{C_{1}+C_{0}}{C_{1}-C_{0}}.
\end{equation}

Next, we need an expression for the force applied to the membrane expressed
through the input variable of memcapacitor such as the applied voltage $V$
(see Eqs. (\ref{Ceq1}) and (\ref{Ceq2})). Since the energy of the plate can
be written as
\begin{equation}  \label{eq28}
U_{ext}(z_{cm})=-\frac {\varepsilon_0 V^2 A}{2(h+z_{cm})},
\end{equation}
where $\varepsilon_0 $ is the vacuum permittivity and $A=d\cdot w$ is the
plate area, the force is given by
\begin{equation}
F(z_{cm})=-\frac {\varepsilon_0 V^2 A}{2(h+z_{cm})^2}.  \label{eqFFads}
\end{equation}

\begin{figure}[tb]
\centering{\includegraphics[width=0.50\columnwidth]{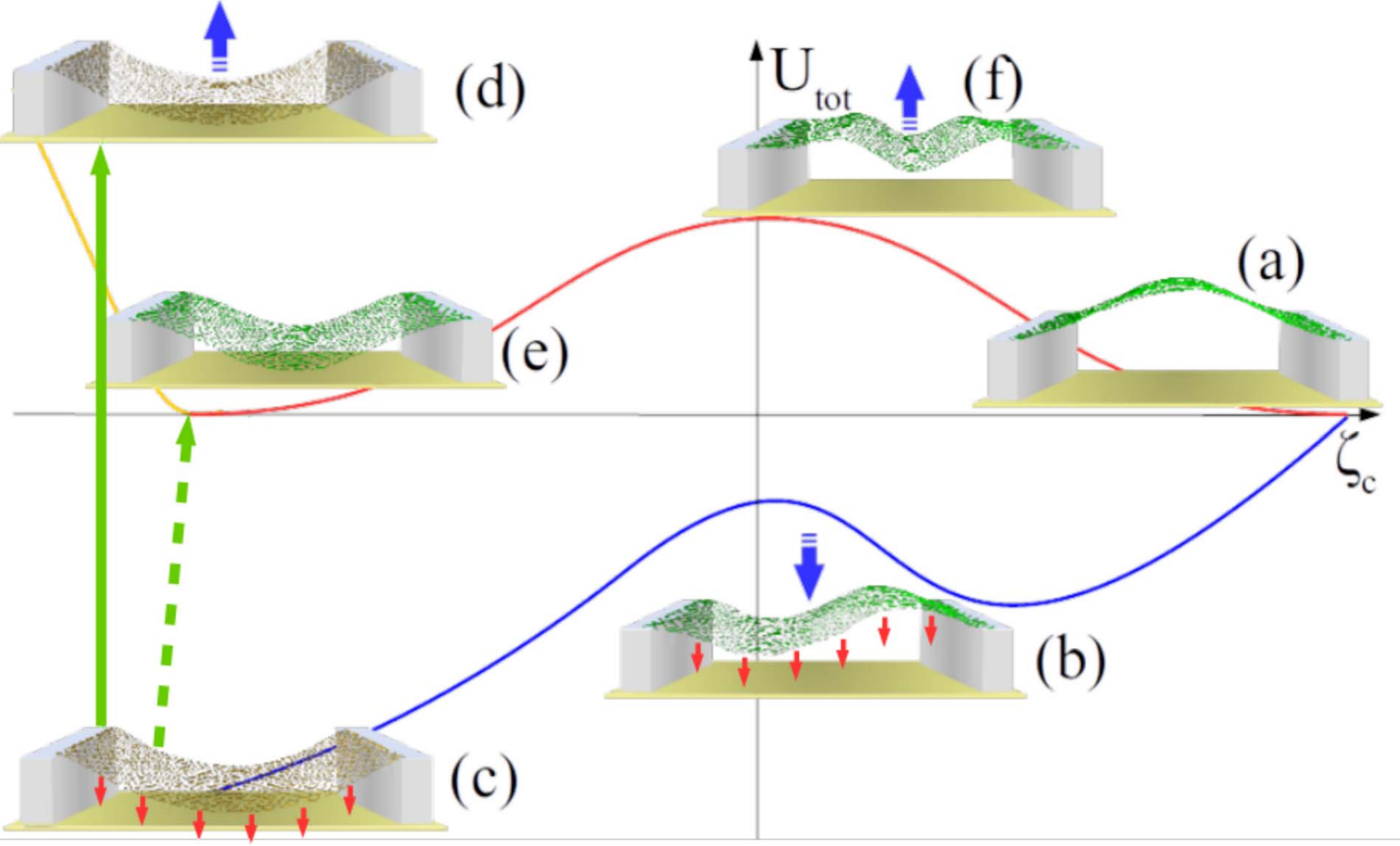}}
\caption{ Schematic representation of transition pathways in the switching
of membrane memcapacitor (red arrows represent the applied electrostatic
force, bold blue arrows represent the movement of membrane). The buckled
upwards membrane (structure (a)) evolves into the buckled downwards state (c)
through a non-symmetric transition state (b). Note that in the presence of force, (c) is
also a transition state. After a slow voltage
removal (dashed green arrow), the membrane remains in the buckled
downwards state (e) (the up-to-down transition). In case of fast force
removal, the membrane does not have enough time to relax its tension (d) and start
moving into the upward state (a) through a symmetric transition state (f) (the
down-to-up transition).}
\label{fig11}
\end{figure}

The overall scheme of memcapacitor switching is schematically presented in
Fig.~\ref{fig11}. Let us consider first the up-to-down transition (path
(a-c,f) in Fig.~\ref{fig11}). As mentioned in  SI Sec.~"Phenomenological Elasticity Theory"
, the smallest threshold force for this transition is associated with the
switching through the non-symmetric state and corresponds to $c_{1}^{\ast
}=0.6948$. In this way, the buckled upwards membrane (structure (a) in Fig.~%
\ref{fig11}) evolves into the buckled downwards state according to (b) and
(c) in Fig.~\ref{fig11}. After the voltage removal, the membrane remains in
the buckled downwards state ((e) in Fig.~\ref{fig11}). In order to estimate
the minimal (threshold) voltage required for this transition, Eq. (\ref%
{eq:fsw_nonsym}) is rewritten accounting for Eq. (\ref{eqFFads}):
\begin{equation}
\frac{\varepsilon _{0}V^{2}dw}{2(h+z_{cm}(c_{1}^{\ast }))^{2}}=263.53\frac{Dw%
}{L^{2}}\sqrt{\frac{L-d}{L}}.  \label{eq:sw_cond_utd_1}
\end{equation}%
Consequently,
\begin{equation}
V_{0\rightarrow 1}=22.96\frac{h+z_{cm}(c_{1}^{\ast })}{L}\sqrt{\frac{D}{
\varepsilon _{0}\cdot d}\sqrt{\frac{L-d}{L}}},  \label{eq:sw_cond_utd}
\end{equation}%
where $z_{cm}(c_{1}^{\ast })=0.3203\sqrt{L(L-d)}$. Eq. (\ref{eq:sw_cond_utd}%
) provides an easy way to find the threshold voltage necessary for the
up-to-down transition of membrane.

In the down-to-up transition, the initial buckled downwards membrane ((e) in
Fig.~\ref{fig11}) is first stretched by the applied force ((c) in Fig.~\ref%
{fig11}). When the force is removed, the membrane state becomes (d). If the
potential energy of (d) is sufficient to overcome the potential barrier
height symmetrically (f), then the membrane may end up in the buckled upwards
state (a). These general processes are partially described by Eqs. (\ref%
{eq:dtu_energ}), (\ref{eq:dtu_force}). Because of the membrane elongation in
(c,d), a refined value of $z_{cm}$ should be used in relevant calculations.

The threshold voltage can be estimated based on the energy conservation (Eq.
(\ref{eq:dtu_energ})) and force equilibrium (Eq. (\ref{eq:dtu_force}))
conditions accounting for $z_{cm}^*$ from Eqs. (\ref{eq:z_cm1}) and (\ref%
{eq:z_cm2}). One can find that
\begin{equation}
\frac {\varepsilon_0 V^2 d\cdot w}{2(h+ z_{cm}^*)^2} = 79.62 \sqrt{2DE_{2D}}
\frac{w}{L} \frac{(L-d)}{(2L-d)}
\end{equation}
and
\begin{equation}
V_{1\rightarrow 0} = 12.62(h+z_{cm}^*)\sqrt{\frac{\sqrt{2 D E_{2D}}}{%
\varepsilon_0\cdot d\cdot L }\frac{(L-d)}{(2L-d)}} .  \label{eq1122}
\end{equation}

An additional aspect of membrane dynamics --
an estimation of thermally-induced switching time -- is considered
in SI Sec. "Stability of buckled membrane".

\section{Concluding Remarks}

\label{sec6}

We have investigated the snap-through transition of a buckled graphene membrane
using a variety of computational and theoretical tools. The main results of
this paper are the expressions for the threshold switching forces (Eq. (\ref%
{F_star}, \ref{eq:fsw_nonsym}) for the up-to-down transition and Eq. (\ref%
{Fup}, \ref{FZCM}) for the down-to-up one) and corresponding voltages (Eqs. (%
\ref{eq:sw_cond_utd}) and (\ref{eq1122})). Our analytical results are
supported by the results of numerical simulations, MD and DFT calculations.
We expect that our findings will find applications in the design,
fabrication and analysis of membrane-based memory devices.

\section*{Acknowledgments}
This work has been supported by the Russian Science Foundation grant No.
15-13-20021. OVI and SNS acknowledge a partial support by the State Fund for
Fundamental Research of Ukraine (grant No. F66/95-2016).

\section{Author contributions statement}

R.D.Y. and Y.V.P. did the MD calculations and developed the \nameref{sec3}.
O.V.I. and S.N.S. developed the \nameref{Sec:Elasticity}. O.V.S. did the DFT calculations.
All authors discussed the results and co-wrote the manuscript.

\section{Additional information}

\textbf{Competing financial interests: }The authors declare no competing
financial interests.

\newpage

\setcounter{secnumdepth}{0}
\def\theequation{S\arabic{equation}}
\def\thefigure{S\arabic{figure}}
\noindent{\huge\bf Supplementary Information}

\section{MD Simulation details}

\label{Sec:MD_Sim_det}

We used NAMD2 (see Ref.~[\citenum{phillips05}]) -- a parallel
classical molecular dynamics package -- to simulate the dynamics of buckled
graphene nanoribbons subjected to external force. NAMD was developed by the
Theoretical and Computational Biophysics Group in the Beckman Institute for
Advanced Science and Technology at the University of Illinois at
Urbana-Champaign.

\begin{figure}[tbh]
	\centering{\includegraphics[width=0.50\columnwidth]{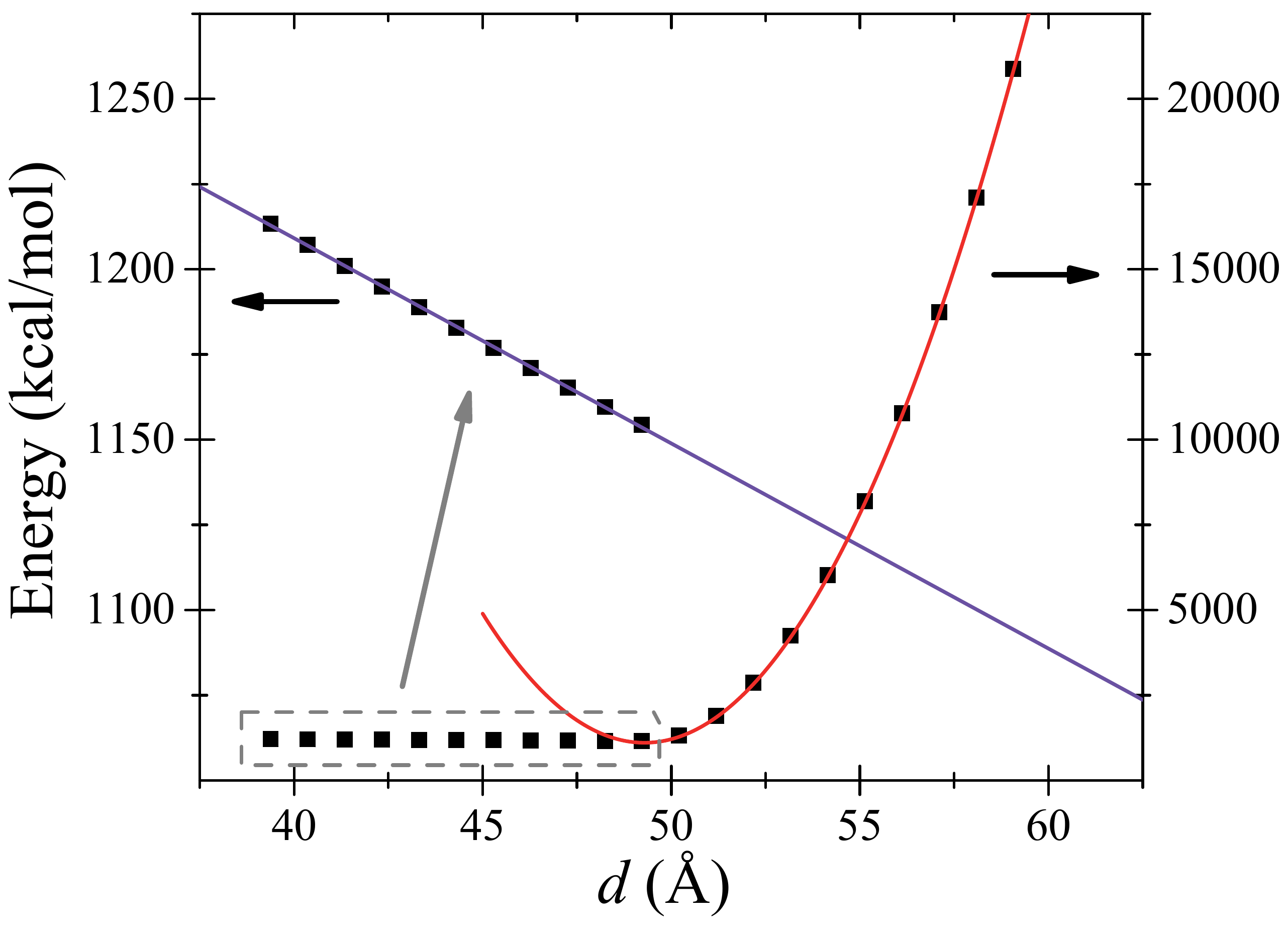}}
	\caption{Calculated energies of a stretched and compressed graphene
		nanoribbon of the free length of $L=49.2$ \AA\ and width of $w=40$ \AA\ as a
		function of the distance $d$ between its fixed edges. This calculation was
		performed to verify the selected force-field parameters of carbon atoms in
		graphene. The fitting curves are represented by lines and are based on $%
		E_{2D}=342$ N/m and $D=1.6$ eV parameters values. The calculation was
		performed using hinged boundary conditions.}
	\label{fig2}
\end{figure}

In graphene, the interactions between the carbon atoms were described using
the standard 2-body spring bond, 3-body angular bond (including the
Urey-Bradley term), 4-body torsion angle and Lennard-Jones potential energy
terms \cite{JCC21367}. The interaction constants were optimized in order to
fit the experimentally observed properties of graphene. The known values
were selected for the equilibrium angles. The Lennard-Jones coefficients
were chosen to match the AB stacking distance and energy of graphite \cite%
{Chen2013}. A global optimization was performed over the remaining
parameters to match the in-plane stiffness ($E_{2D}=342$ N/m), bending
rigidity ($D=1.6$ eV) and equilibrium bond length ($a=1.421$ \AA ) of
graphene. We performed a series of test calculations that have demonstrated
that the in-plane stiffness and bending rigidity of a selected graphene
nanoribbon are in perfect agreement with the listed above values (see Fig. %
\ref{fig2}).

A $1$ fs time step was used and the system temperature is kept at room
temperature with a Langevin damping parameter of $\gamma =0.2$ ps$^{-1}$ in
the equations of motion. The van der Waals interactions were gradually cut
off starting at 10 \AA\ from the atom until reaching zero interaction 12
\AA\ away. In each simulation, the buckled membrane was used as an initial
condition. Typically, the system was equilibrated for 100 ps in the presence
of the applied force and its subsequent dynamics (after the force removal)
was typically simulated for 100 ps (these times were changed to 500 ps
in the case of overdamped simulations). The final state was determined according to
$z$-coordinate of an atom in the cental part of the membrane.

\begin{figure}[tbh]
	\centering{\includegraphics[width=0.50\columnwidth]{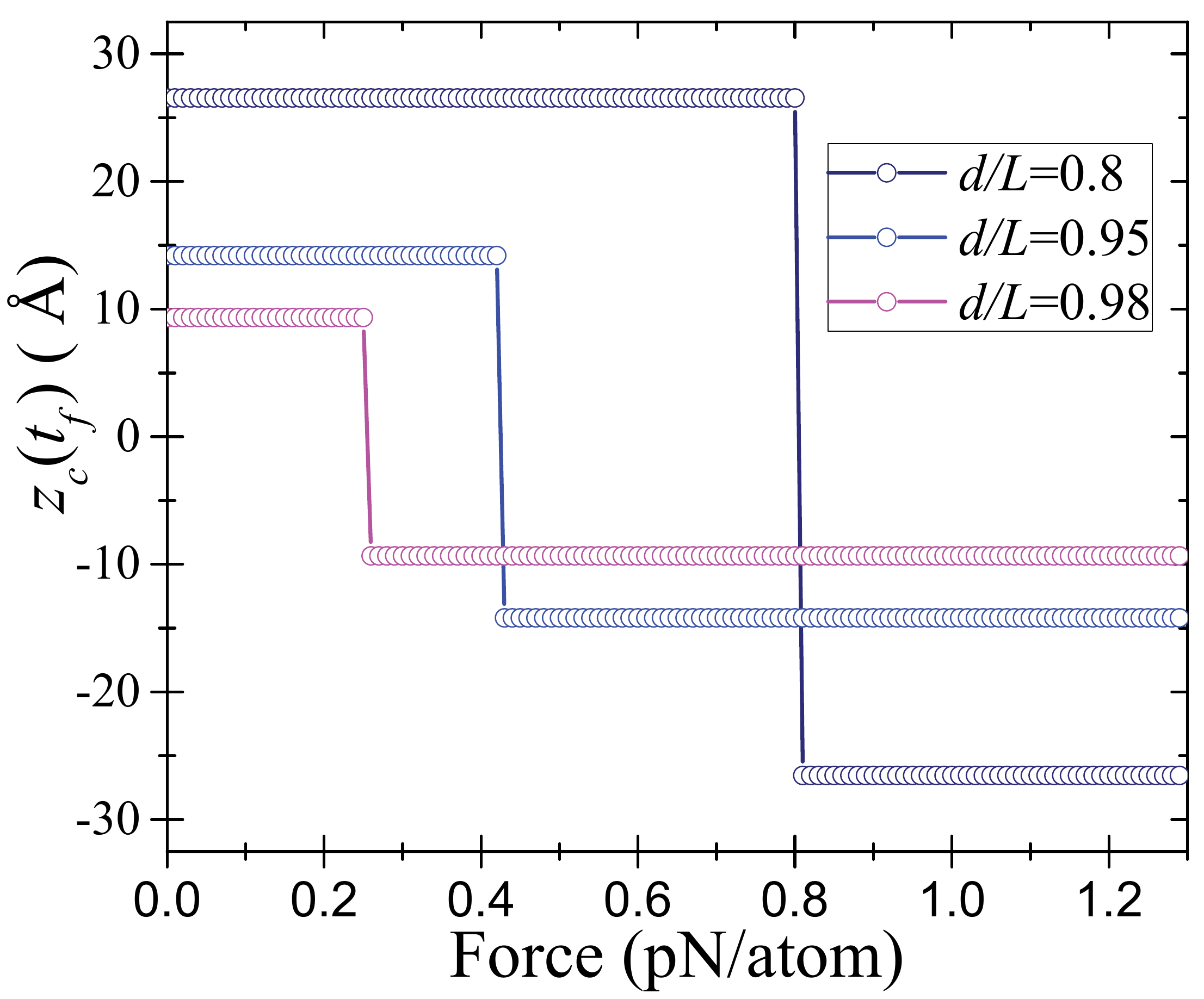}}
	\caption{Up-to-down transition: the final position of a central atom of membrane as a function
    of the applied force magnitude in an overdamped calculation ($T=0$, $\protect\gamma =1$ ps$^{-1}$)
    for membrane \textbf{\textit{B}} (42 rings length).}
	\label{fig5}
\end{figure}

We also performed a series of overdamped simulations of membrane dynamics.
This type of simulations was made using zero temperature and strong damping
that significantly suppress the kinetic energy of the membrane. As a result,
we have obtained more regular final states of the membrane (see Fig. \ref%
{fig5}) and slightly different values of the threshold switching force,
which, however, can be more easily interpreted as now the kinetic energy
can be disregarded. Comparing Fig. 2(b) and
Fig. \ref{fig5} we note that the overdamped
calculations provide somewhat larger/comparable estimations for the
threshold switching force.

\section{Elasticity theory}

\subsection{Down-to-up transition}

\label{Sec:Down-to-Up-Appendix}\

\begin{figure}[tbh]
	\centering{\includegraphics[width=0.49\columnwidth]{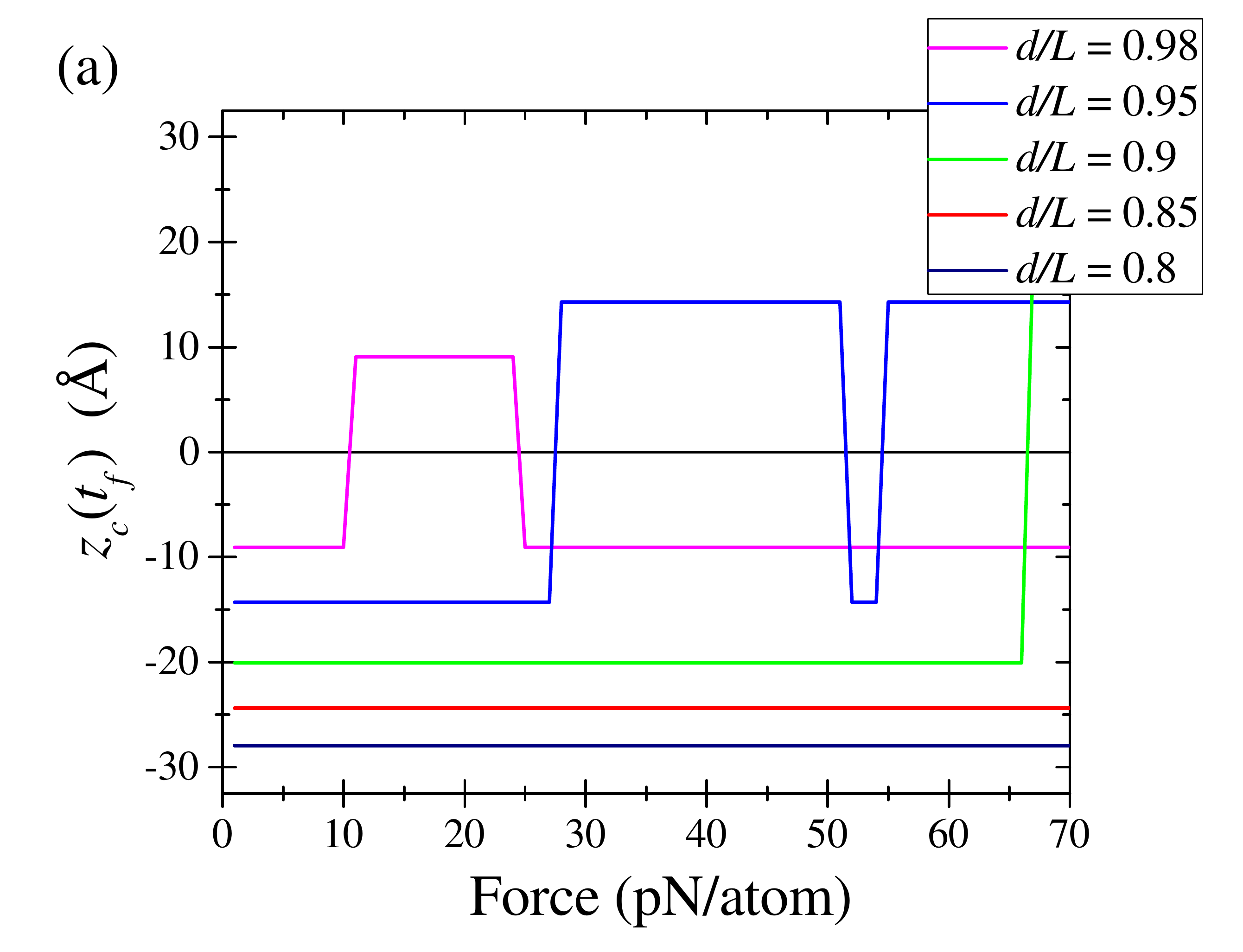}
	\includegraphics[width=0.49\columnwidth]{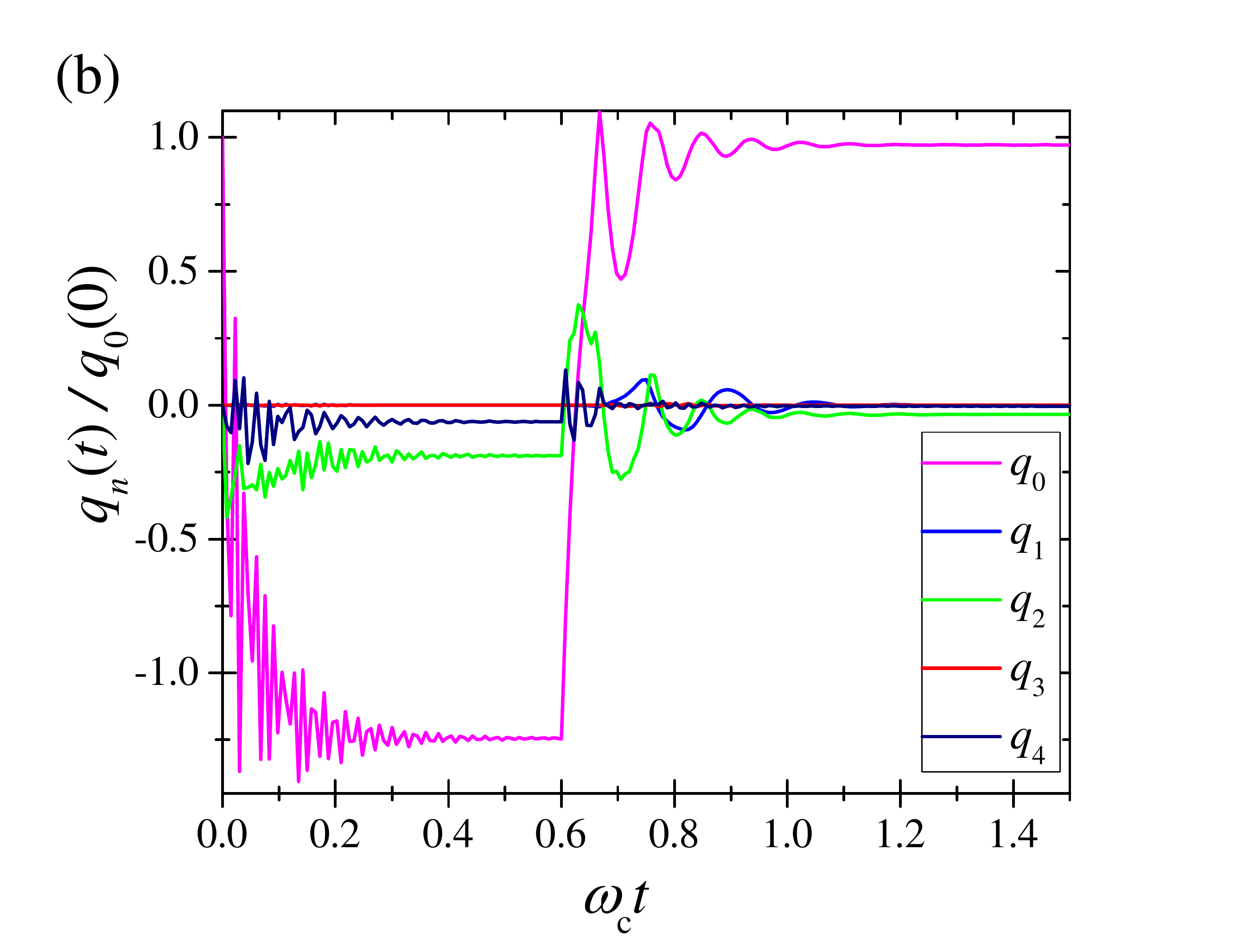}}
	\caption{Numerical simulation of the down-to-up transition. (a) The
		central-point deflection $\protect\zeta _{\mathrm{c}}$ versus the applied
		transverse force $F$ for several values of $d/L$. Snapping-through appears
		at the critical force value $F=F^{\uparrow }$. (b) Time-dependence of the
		first few harmonics amplitudes $q_{n}(t)$.}
	\label{Fig:iii}
\end{figure}

A study of the down-to-up transition based on the standard elasticity theory is presented in Fig.~\ref{Fig:iii}.
Fig.~\ref{Fig:iii}(a) shows the final position of membrane depending on the applied force
magnitude. In these calculations, the force was applied for a finite interval of time and removed at $\omega_ct=0.6$.
It was observed that the final state of membrane depends on few modeling parameters
(the applied force, damping rate, etc.). The possibility of the down-to-up transition
is clearly seen in Fig.~\ref{Fig:iii}(a). The time-dependence of  harmonics amplitudes is
exemplified in Fig.~\ref{Fig:iii}(b) for a particular value of force.
 We note that in Fig.~\ref{Fig:iii}(b) the
amplitudes $q_{n}$ are normalized to $q_{0}$ at $t=0$.
Fig.~\ref{Fig:iii}(b) shows that the harmonics amplitudes are modified by the
applied force (e.g., it is clearly seen that $|q_{0}(t)/q_{0}(0)|>1$ at $\omega_ct<0.6$).
Importantly, the amplitudes of higher harmonics are small and can be neglected.

\clearpage
\subsection{Phenomenological Elasticity Theory}

\label{Sec:Elast_Phenom}\

The Cartesian coordinates, which give the position of a membrane element defined by the internal coordinate $s$, are related to the angle $\theta_{i}$ (see Eqs. (16) and (17)) as
\begin{eqnarray}
x_{i}(s) &=&L\int\limits_{-1/2}^{s}\cos \theta _{i}(s^{\prime })\text{d}%
s^{\prime },  \label{eq11} \\
\zeta_{i}(s) &=&L\int\limits_{-1/2}^{s}\sin \theta _{i}(s^{\prime })\text{d}%
s^{\prime }.  \label{eq12}
\end{eqnarray}%
According to the above equations, $x_{i}(-1/2)=0$ and $\zeta_{i}(-1/2)=0$. Another pair
of boundary conditions is given by
\begin{equation}
x_{i}(1/2)=d,\qquad \qquad \zeta_{i}(1/2)=0.  \label{eq:constraints}
\end{equation}%
In the limit of small deflections, the boundary conditions, Eq. (\ref{eq:constraints}), can be resolved for the coefficients $A_{i}$ and $c_{i}$ in Eqs. (16) and (17):
\begin{eqnarray}
A_{\text{s}} &=&\pm \frac{48\sqrt{385}\sqrt{1-\frac{d}{L}}}{\sqrt{
		5-88c_{0}^{2}+528c_{0}^{4}}}, \\
c_{2} &=&-\frac{1}{20}c_{1}, \\
A_{\text{ns}} &=&\pm \frac{c_{1}240\sqrt{35}\sqrt{1-\frac{d}{L}}}{\sqrt{
		1200c_{1}^{4}-56c_{1}^{2}+3}}.
\end{eqnarray}%
Therefore, the value of a single parameter $c_{0}$ (or $c_{1}$) defines
completely the symmetric (or non-symmetric) shape of membrane.

\begin{figure}[tb]
	\centering{\includegraphics[width=0.50\columnwidth]{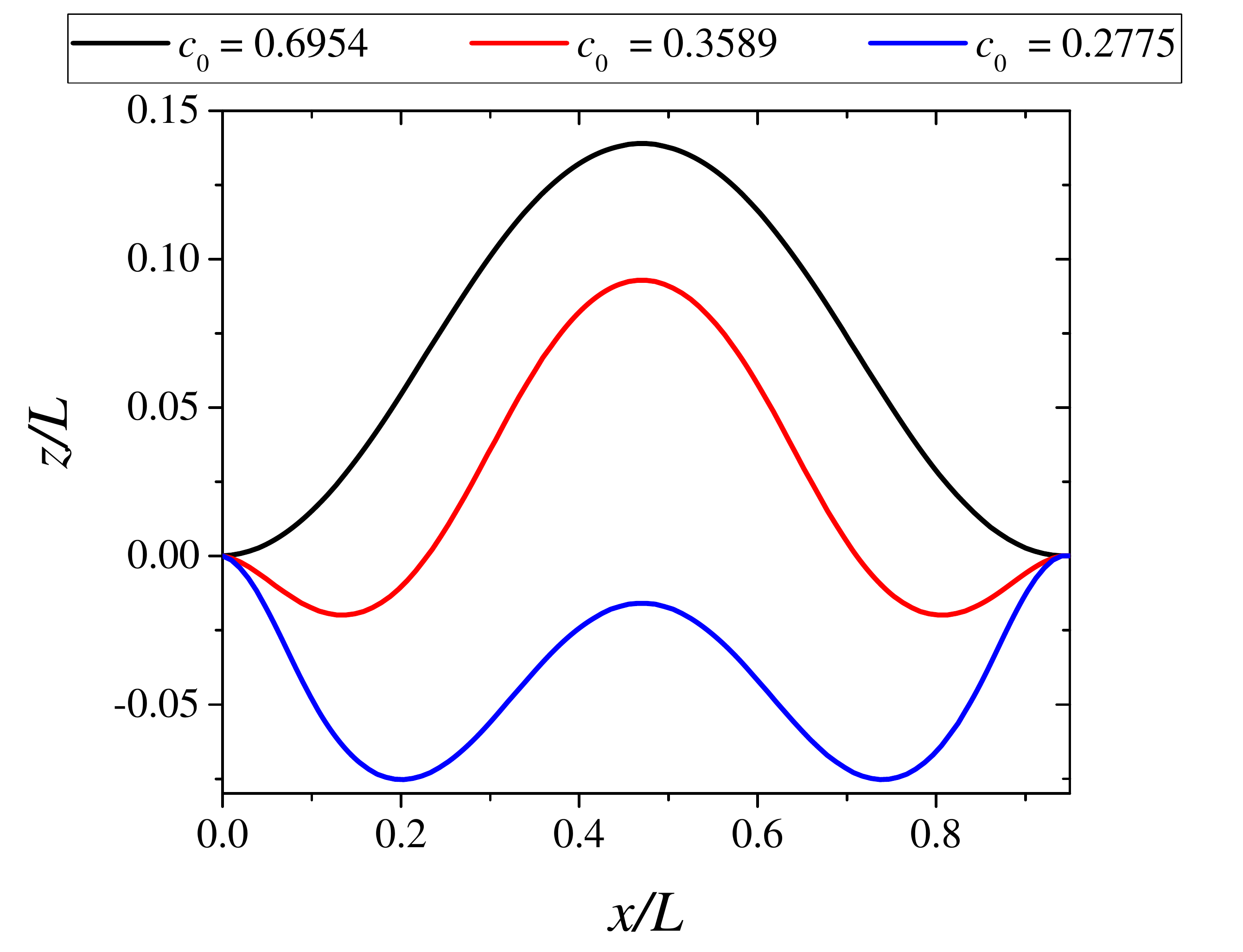}}
	\caption{Symmetric shapes of membrane calculated at $d/L=0.95$. The curves
		from the top to bottom: the geometry minimizing the bending energy, geometry
		at the threshold switching force, and geometry maximizing the bending
		energy. }
	\label{fig7}
\end{figure}

In the case of the symmetric shape of membrane (Eq. (16)), at $%
F=0$, there are two possible solutions of Eq. (19):
\begin{equation}
c_{0,1} = \sqrt{\frac{\sqrt{5}}{11}+\frac{37}{132}} \approx 0.6954
\end{equation}
and
\begin{equation}
c_{0,2} = \frac{1}{2} \sqrt{\frac{1}{33} \left(37-12 \sqrt{5}\right)} \approx
0.2775  \label{eq787}
\end{equation}
valid for any $d/L$. These values correspond to the bending energies $%
U_{b,\text{s},1(2)}=k_{1(2)}Dw(L-d)/L^2$ with $k_1=39.5$ (minimum of $U_{b,\text{s}}$) and
$k_2=200.5$ (maximum of $U_{b,\text{s}}$). The corresponding geometries are shown
in Fig. \ref{fig7}. Importantly, the snap-through transition of membrane
takes place as $c_0$ changes from $0.6954$ to $0.2775$. We note that Eq. (19) can be used to find $F_\text{s}$ needed to 'support' the membrane
profile defined by a given value of $c_0$.

\begin{figure}[tb]
	\centering{\includegraphics[width=0.50\columnwidth]{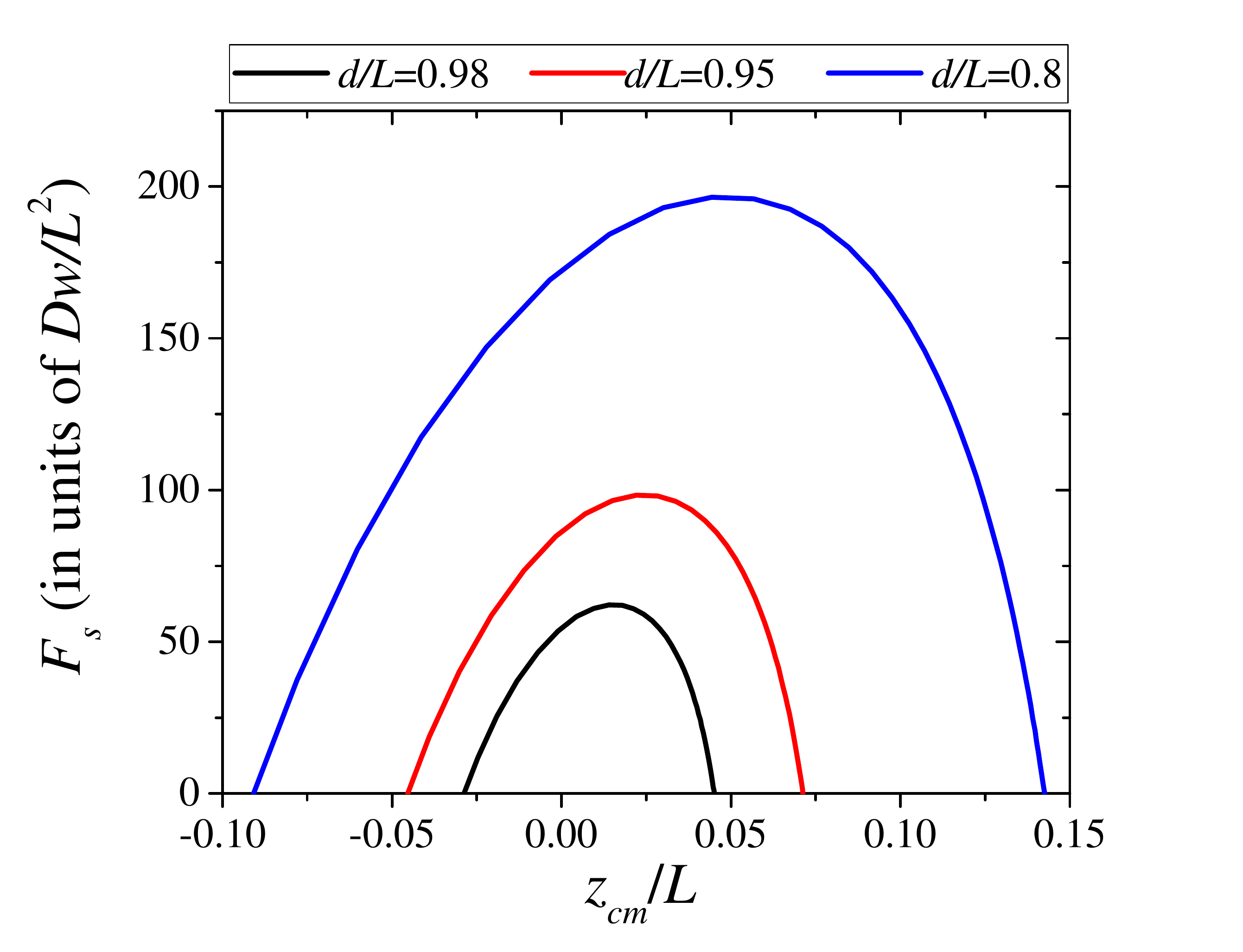}}
	\caption{The applied force $F_{\text{s}}$ as a function of $z_{cm}$ found for the
		symmetric membrane profile in the adiabatic switching regime. The maximum of
		$F_{s}$ corresponds to the threshold switching force through the symmetric
		membrane profile.}
	\label{fig8}
\end{figure}

The force as a function of $z _{cm}$ is presented in Fig. \ref%
{fig8} for several values of $d/L$.

\subsubsection{Dynamic regime}

\label{sec3c}

Here we assume that the membrane transition is induced by an abruptly
applied force (similar to the case of our MD simulations). Under the
assumption of energy conservation, the threshold switching force can be
found from the condition that the initial energy equals the potential energy
barrier height (that depends on the applied force).

Technically, the calculation can be performed as follows. The value of
parameter $c_{i}$, $c_{i}^{\ast }(F)$, corresponding to the maximum of $%
U=U_{b}+U_{ext}$ (in the presence of $F$, see Eq. (18)) can
be obtained from $\text{d}U/\text{d}c_{i}=0$. Next, $c_{i}^{\ast }(F)$ is
substituted into
\begin{equation}
U_{b}\left( c_{i}^{0}\right) +F\zeta \left( c_{i}^{0}\right) =U_{b}\left(
c_{i}^{\ast }\right) +F\zeta \left( c_{i}^{\ast }\right) ,  \label{eq344}
\end{equation}%
where $c_{i}^{0}$ corresponds to the potential energy minimum at $F=0$ (note
that $U_{b}(c_{i}^{0})$ is the energy corresponding to $k_{1}$ given below
Eq. (\ref{eq787})). Finally, the threshold switching force is found from Eq.
(\ref{eq344}).

If one assumes that, at $t=0$, the membrane has the non-symmetric profile,
then one can get that the dynamic threshold switching force is the same as $%
F^{\downarrow}_{\text{ns}}$ given by Eq. (21). This result, however, should only be
used as an upper estimate. Indeed, initially, the membrane has the symmetric
profile and determining the exact point of switching from the symmetric to
non-symmetric shape is a challenge.

\begin{figure}[tb]
	\centering{\includegraphics[width=0.50\columnwidth]{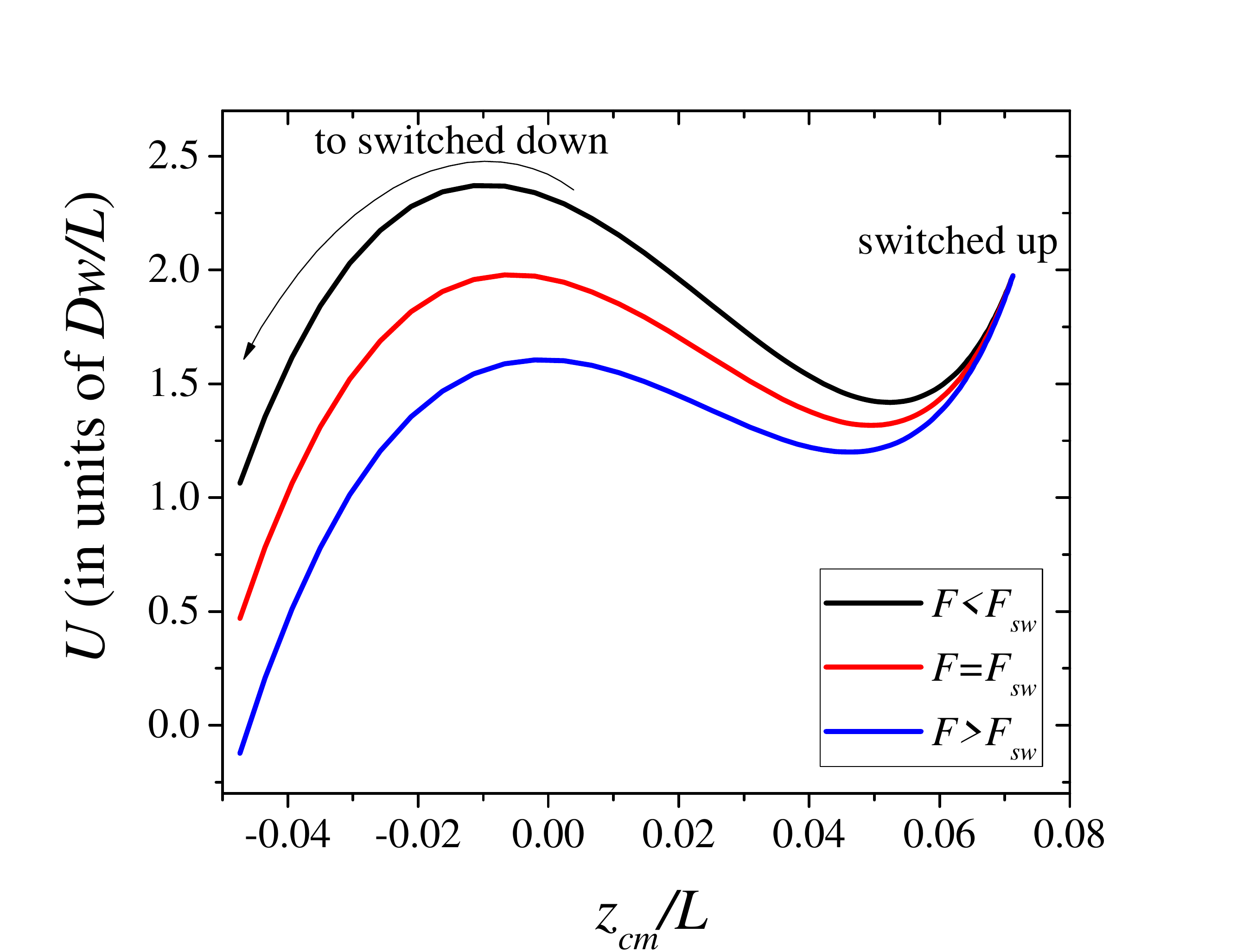}}
	\caption{Potential energy landscape for the symmetric switching at several
		values of the applied force and $d/L=0.95$.}
	\label{fig10}
\end{figure}

Unfortunately, the above mentioned equations can not be solved analytically
for the transition through the symmetric profile. Figure \ref{fig10} shows the
numerically found potential energy as a function of position of the center of mass
for several representative values of the applied force. This plot
demonstrates that the potential barrier height becomes lower than the
initial energy when the applied force becomes larger than the threshold
switching force $F_{sw}$.

Various threshold switching forces for the up-to-down transitions are
summarized in Fig.~8 together with the results of our MD and DFT calculations.
This plot shows that $F^{\downarrow}$ provides the best phenomenological approach estimation for our
MD results. The deviation of MD points from $F^{\downarrow }$ curve at
smaller $d/L$ can be related to the approximation of small deflections used
in our analytical model. Additionally, MD simulations involve energy
relaxation channels not included into the simplified
model given by Eq. (\ref{eq344}).
Overall, however, we are quite satisfied with the results of our
phenomenological model.

\section{DFT Calculations} \label{Sec:DFT}

\subsection{Computational details}

Our DFT calculations were carried out using the Jaguar quantum-chemistry \cite{Jaguar} package within the electron density functional approach with the
use of a Becke three-parameter hybrid functional \cite{Becke93}, a
Lee--Yung--Parr correlation functional \cite{Lee88} (the B3LYP method), and
a 6-31G basis set of atomic orbitals. Optimization of structures was
performed by an analytical method up to a gradient of the displacement of
atomic positions of $10^{-4}$ au. The minimum on the potential energy
surface was identified by the absence of imaginary values in the matrix of
second derivatives. As a test calculation, we have identified that the stretching of an armchair-edged graphene nanoribbon
($L=20.9$ \AA, $w=14.7$ \AA) yields the in-plane
stiffness $E_{2D}^{DFT}=338$ N/m, which is in good agreement with the experimental
value $E_{2D}=340$ N/m.\cite{Lee08}

We investigated a zigzag graphene nanoribbon of the length of $L=24.6$
\AA\ (10 rings) and of the width of $w=11.4$~\AA. The dangling bonds at nanoribbon
boundary were terminated by hydrogen atoms.
The hinged boundary conditions were realized by fixing the  armchair edges
at all calculation steps. The nanoribbon D1 with clamped boundary
conditions was constructed from the flattest configuration C1 (with hinged boundary conditions) by an elongation of
each armchair side by two rings kept fixed during
calculations.

\subsection{Up-to-down transition}

First of all, we note that the optimization of unloaded configurations C1, C2, and C3 gives the bending rigidity
$D^{DFT}=1.75$ eV, which is in good agreement with the experimental value $%
D=1.6$ eV~\cite{Zhang11}.

\begin{figure}[t]
	\centering{\includegraphics[width=0.40\columnwidth]{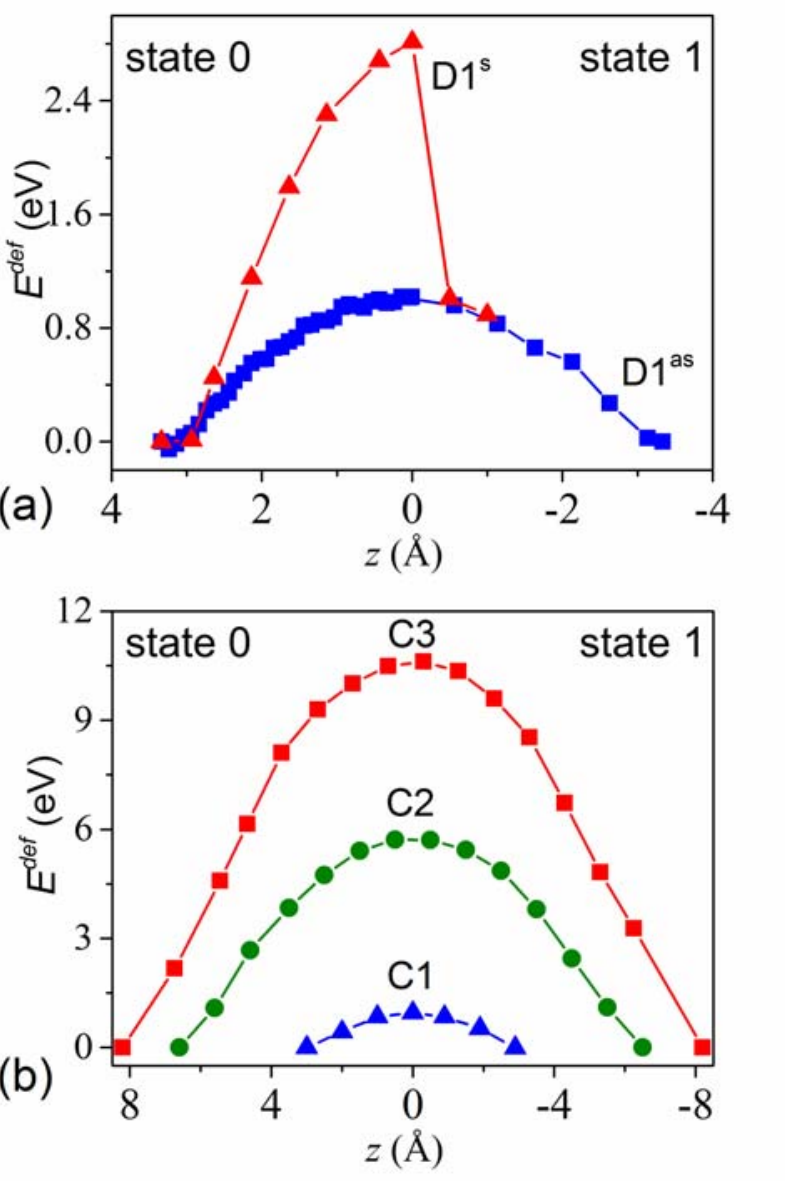}}
	\caption{DFT deformation energy versus the position of central atoms in the process of the up-to-down switching. Based on DFT approach, we were able to identify
		(a) both symmetric and non-symmetric paths for the case of clamped boundary conditions, and (b) only the non-symmetric path for the case of hinged boundary conditions.}
	\label{Fig:DFT2}
\end{figure}

As reported in Ref.~[\citenum{OSedelnikova2016}],
while, the non-symmetric path of switching is preferable due to the lower energy barrier,
at small deflections from the unloaded state, the nanoribbon tends to be symmetric.
A detailed investigation of the up-to-down transition of D1 shows that at small deformations, the
difference between the energies of the symmetric and assymetric shapes is small
(for instance, the initial displacement of membrane by $0.5$~\AA~requires about $0.02$ eV and $0.05$ eV for the
symmetric and non-symmetric paths, respectively, see Fig.~\ref{Fig:DFT2}(a)). However, the energy difference between the paths increases significantly with deformation.
When the positions of central atoms are close to the level of fixed edges, the deformation energy has a maximum, which was used to estimate
the up-to-down threshold switching force~\cite{OSedelnikova2016}. We note that the symmetric path in Fig.~\ref{Fig:DFT2}(a) was generated by an increase of the $z$-coordinate of central atoms starting at $z\sim 0$.

The deformation energy of buckled graphene
membrane with hinged boundary conditions is plotted in Fig.~\ref{Fig:DFT2}(b). In the case of hinged boundary conditions,
only the non-symmetric path  has been identified in the framework of
our DFT approach. It was found that the transition energy barrier is smaller for larger values of $d/L$.

Overall, the up-to-down threshold switching force estimated from DFT calculations agrees with both MD simulation and elasticity theory results (see Fig.~8). This plot presents the threshold switching force in the units of $Dw/L^2$ for C1 and C2 configurations. We note that the DFT value for the threshold force for C3 is about $235.3 Dw/L^2$. As $d/L$ of C3 is significantly less than 1, the DFT threshold switching force for C3 is not the right quantity to compare with results of the linear elasticity theory.

\section{Stability of buckled membrane}

The information storage time in the state of buckled membrane is limited by the finite temperature which leads
to thermally-activated switching between two stable membrane states at long times. In the classical approach~\cite{Atkins}, the rate constant $k$ of thermally-activated switching can be approximately described by Arrhenius equation
\begin{equation}
k=\mathcal{A} \cdot e^{-\frac{E_b}{k_B T}}, \label{eq:k}
\end{equation}
where $\mathcal{A}$ is the pre-exponential factor (according to some more advanced theories~\cite{Eyring35a,Hanggp90a}, it describes the energy distribution in the system) and $E_b$ is the potential barrier height. In the case of buckled membrane, $E_b$ corresponds to the energy of
the intermediate asymmetric state of membrane during its switching (the saddle point state) with respect to the energy of its stable states.

Next we make an estimation for the thermally-activated switching time of $d/L=0.94$ membrane of type B (a detailed study of the membrane stability is beyond the scope of this paper). For this purpose we first performed a series of MD simulations of the dynamics of $d/L=0.98$ membrane without applied force. These simulations have indicated that $d/L=0.98$ membrane stays in its initial state for at least 10 ns without switching (this is, of course, an underestimate of its storage time). Then, assuming that the change of the pre-exponential factor in Eq. (\ref{eq:k}) with $d/L$ compared to the corresponding change in the exponential factor is not significant, and using $E_{0.98}=0.60$ eV, $E_{0.94}=1.8$ eV, and $T=300$ K one finds $k^{-1} > 40,000$ years. Clearly, such  stability is sufficient for practical applications.

\end{document}